\documentclass[12pt,preprint]{aastex}
\usepackage{graphicx}
\newcommand{\mh}{H$_2$ }
\newcommand{\mha} {H$_2$}

\newcommand{\mj}{M$_{J}$}
\newcommand{\s}{$\sim$}
\newcommand{\bd}{\begin{displaymath}}
\newcommand{\ed}{\end{displaymath}}
\begin{document}

\title{Models of Chemistry, Thermal Balance, and Infrared Spectra from 
Intermediate-Aged Disks around G and K Stars}

\author{U.Gorti}
\affil{University of California, Berkeley, CA}
\author{D.Hollenbach}
\affil{NASA Ames Research Center, Moffett Field, CA}

\begin{abstract}
We model gas and dust emission from regions 0.3$-$20 AU from a
central low mass star in intermediate-aged (\s $10^7$ years) disks whose
dust is fairly optically thin to stellar radiation.
The models treat thermal balance and chemistry self-consistently, 
and calculate the vertical density
and temperature structure of the gas in a disk. The gas and
dust temperatures are calculated separately.
The models cover gas masses $10^{-3}-
1$ \mj, dust masses  $10^{-7} - 10^{-4}$ \mj, and
treat solar type (G and K) stars. We focus on mid-infrared
and far-infrared emission lines from various gas species such as
the rotational lines of \mha, OH, H$_2$O and CO molecules and
the fine structure lines of  carbon, oxygen, sulfur, iron, and silicon 
atoms and ions.
These lines and the dust continuum are observable by the Spitzer Space
Telescope, and future missions including SOFIA and the Herschel
Space Observatory. We
find that the [SI]25.23$\mu$m line is the strongest emission line
 for a wide range of disk and stellar parameters, followed
by emission from [SiII]34.8$\mu$m, [FeII]26$\mu$m,
and [OI]63$\mu$m. [FeI]24$\mu$m and rotational lines of OH and
H$_2$O are strong when gas masses
are high ($\gtrsim$ 0.1 \mj).  Emission from the
rotational lines of \mh is more difficult to detect, unless disk gas 
masses are substantial ($\gtrsim $ 0.1 \mj). 
For emission from \mh lines to be observable and yet the dust be optically
thin in stellar light,  the ratio of gas to small
sub-millimeter sized dust particle mass in the disk needs to be
$\gtrsim 1000$, or at 
least an order of magnitude higher than that in the 
interstellar medium. This may be possible at intermediate stages in
disk evolution, such as in the gas gathering stage of the core
accretion scenario for giant planet formation,
 where most of the dust has coagulated into larger 
objects ($\gg 1$mm) but the gas has not yet fully dispersed. Whereas
the absolute fluxes observed in some lines such as [FeI]24$\mu$m and
\mh S(0) 28$\mu$m primarily measure the gas mass
in the disks, various line ratios diagnose the inner radius of the gas,
and the radial temperature and surface density distribution of the gas.
Many lines become optically thick and/or suffer chemical or thermal effects 
such that the line luminosities do not increase appreciably
with increasing gas mass. We predict it may be difficult for the
Spitzer Space Telescope to detect $\lesssim 1 $ M$_J$ of gas 
in optically thin (in dust) disks
at distances $\gtrsim$ 150 pc. The models presented here will be useful
in future infrared studies of the timescale for the dispersion of gas
in a planet-forming disk, and testing core accretion models of
giant planet formation. 
\end{abstract}
  

\section{Introduction}
Circumstellar disks evolve from gas-rich structures with  1\% of the 
total mass in dust to disks which are observed to be completely devoid of gas 
and dominated by emission from dust particles. During this process
of evolution, the disk, which is initially very optically thick in dust
with radial midplane optical depths in the visual $\tau_V \gtrsim 10^3$,
becomes optically thin with $\tau_V$\s 1  in
a few million years (Haisch, Lada \& Lada 2001). At this intermediate stage 
of disk evolution planetary formation is probably underway with dust
particles colliding and growing to form larger objects (which greatly reduces $\tau_V$),
and with  some residual gas.  Disks with ages $\gtrsim 10^7$ years 
are observed to have almost no gas (Zuckerman et al. 1995; Duvert et al. 2000). 
Smaller dust particles in these disks have either dispersed or have coagulated
to larger sizes, and these
disks are very optically thin in dust with $\tau_V \lesssim 10^{-2}$. 
These very optically thin disks are called ``debris disks'', as
the dust in these disks is generally not primordial but is
continually generated debris from colliding rocky
bodies  and planetesimals. 

The evolution of the dust component of disks has been extensively studied
(e.g., Weidenschilling 1977; Weidenschilling et al. 1997)
in models where the dust grains in disks grow in size via collisional
agglomeration. As dust grains grow and planetesimals eventually form, 
the disk becomes optically
thin to stellar radiation and the fractional dust luminosity, $f_D
= L_{IR}/L_{bol}$, the ratio of the disk infrared luminosity to the stellar bolometric
luminosity,  is observed to decrease with the age of the disk as
$t^{-1.76}$ (Habing et al. 1999; Spangler et al. 2001; Zuckerman 2001).
This decrease in dust luminosity occurs as micron-sized or smaller 
particles are lost through the effects of radiation pressure, 
Poynting-Robertson drag and collisions (e.g., Backman \& Paresce 1993), and the
population of larger bodies which are the ultimate source of the dust dwindles.

Gas evolution in disks is less well understood.  CO observational
studies seem to indicate that gas beyond about 30 AU  disappears rather abruptly at disk ages 
$\gtrsim 10^7$ years (e.g., Zuckerman et al. 1995; Duvert et al. 2000). 
Theoretical studies also indicate short gas dispersal timescales (Hollenbach,
Yorke \& Johnstone 2000;
Clark, Gendrin \& Sotomayor 2001). However, there is some controversy regarding the
presence or absence of gas in disks of ages $\sim 10^7$ years. 
Seemingly contradicting the CO observations,
 there may have been detection of \s $0.1 - 1$ \mj \ of
molecular hydrogen gas in three younger (ages $\gtrsim 10^7$ years) debris
disks by Thi et al. (2001) using ISO. These ISO results are controversial
(see LeCavalier des Etangs et al. 2001; Richter et al. 2002; Sheret et al. 2003;
Brandeker et al. 2004). If the ISO
observations are confirmed, then perhaps CO is
not a reliable tracer of gas mass in such systems, and may be subject
to chemical effects such as freezing onto cold dust grain surfaces in the outer disks and
optical depth effects in the inner disk. On theoretical grounds, although
gas in the outer regions ($r \gtrsim 10-100$ AU) may be subject to photoevaporation 
due to stellar radiation, gas inside these radii does not photoevaporate because
it is more tightly bound gravitationally to the star. Viscous accretion onto the central
star and viscous spreading to the outer photoevaporating disk likely dominates the
gas dispersal of the inner ($r \lesssim 10-100$ AU) disk (Johnstone et al. 1998;
Hollenbach et al. 2000; Clarke et al. 2001). However, the timescale for viscous dispersal at
$r \sim 10-100$ AU is uncertain. 

A determination of gas dispersal timescales in disks is important for 
understanding the formation of planetary systems. Two competing theories
exist for the formation of gas giants. Gaseous Jovian-type planets are
believed to form in disks via accretion of gas onto rocky cores of a few
earth masses (e.g., Lissauer  1993; Pollack et al. 1996;
Kornet, Bodenheimer \& Rozyczka 2002) or by gravitational instability 
in disks leading to the formation of clumps which subsequently contract 
to form giant planets (e.g., Boss 2003). The presence or absence of gas in 
intermediate-aged or transitional disks ($\tau_V \sim 1$, ages \s $10^7$ years)
can potentially discriminate between these two theories. In the gravitational
instability scenario, gas giants form quickly and the gas may dissipate 
rapidly. Longer gas disk lifetimes($\gtrsim 10^6$ years) facilitate the 
formation of gas giant planets through conventional core accretion models,
allowing sufficient time 
for the building of a rock/ice core of a few $M_{\oplus}$ 
and subsequent accretion of gas. At the gas accretion  stage, 
there would be little dust (or equivalently, particles small enough to 
significantly radiate at infrared/sub-millimeter wavelengths and 
be observed) in the gas giant regions of disks, 
and substantial amounts ($\gtrsim$ 1 \mj, or a Jupiter mass) of gas. 
Core accretion theory suggests that systems may reside in this state
for a significant length of time (few Myr, e.g., Pollack et al. 1996).
At this stage, the ratio of gas mass to dust mass (where dust is defined as particles with
sizes $\lesssim$ 1 mm) would initially increase from the interstellar value of 100 before
the final gas dispersal reduces the ratio to values $\ll$ 100 characteristic of the
older debris disks. The detection of gas infrared emission during this epoch of gas accretion
onto rock/ice cores would help support the core accretion scenario. 
 Conversely, if very little gas ($\ll $1  M$_J$) is present
at the moment when the dust disk first becomes optically thin (signalling the 
buildup of several earth mass cores), then the core accretion model
is likely invalid. 

In the terrestrial planet region of the disk, the residual gas content 
of the disk at the epoch when embryos or protoplanets assemble to form
terrestrial planets helps determine the ultimate mass and 
eccentricity of the planet, and therefore the eventual
habitability of the planet (Agnor \& Ward 2002; Kominami \& Ida 2002;
Kominami \& Ida 2004).
In particular, a finite range of gas masses produces earth-mass planets
on circular orbits. If the gas mass inside 3 AU at ages of $10^{6.5}$ to $10^7$
years is $\gg 10^{-2}$ M$_J$, the tidal interaction of the planets with the
gas is sufficient to circularise the orbits of lunar-mass protoplanets, making it
difficult for collisions between them to build earth-mass planets. 
On the other hand, if the gas inside 3 AU is $\ll  10^{-2}$ M$_J$, earth-mass
planets can be produced, but in orbits considerably more eccentric than that of the Earth. 

At later debris disk stages, $ t \gg 10^7$ years, 
the presence of even small amounts of gas (\s $10^{-2}$ M$_J$)  can affect dust dynamics
because of the drag forces they induce and, therefore, this gas 
 may affect the disk structure significantly.
This effect weakens  the interpretations
of gaps and ring signatures observed in dusty disks as being due to the presence
of planets, as the presence of gas could give rise to the same structures
(Klahr \& Lin 2001; Takeuchi \& Artymowicz 2001;
 Takeuchi \& Lin 2002). 
 
In the work presented here, we  model gas
emission from a disk that has optically thin
or marginally optically thick ($\tau_V \lesssim 5$)
 dust,\footnote{Henceforth, we shall call our models ``optically thin'' even though
as discussed in \S 2.1, they apply for $\tau_V \lesssim 5$, where
$\tau_V$ is the radial optical depth through the midplane.} so is perhaps 
at a stage where  considerable rock/ice planetesimal and 
planet formation has already taken place, but may still retain 
appreciable amounts (from \s $10^{-3}$  M$_J$ to $\gtrsim$ 1 M$_J$)
of gas left over from the rock/ice planet building phase. 
We predict expected emission from
such a range  of gas masses in optically
thin dust disks as a function of the gas mass, dust mass, and stellar 
properties. The results  will aid future observational programs that plan to look for gas emission
in  disks in order to understand the evolution of gas in disks, the timescales for
gas dispersal, and the process of giant planet formation. 
New observational facilities such as the Spitzer Space Telescope, SOFIA and the 
Herschel Space Observatory
are sensitive enough to detect infrared and
sub-millimeter emission from small amounts of gas using tracers
other than CO. These IR tracers may be better suited than CO for
detecting gas in the 1 AU $-$ 20 AU planet forming region around solar-type
stars.  The possibility of detection of gas emission from 
the planet-forming regions around intermediate-aged and evolved
debris disks  by future IR 
observational studies is one of the main motivations of the present paper. 

There has been
extensive work on disk modeling of both young optically thick disks and older
optically thin disks. However, most of this work has concentrated on either
solving the disk structure by setting the gas and dust temperatures to be equal
(e.g. D'Alessio et al. 1998, 1999; Dullemond 2002; Dullemond,
van Zadelhoff \& Natta 2002; Lauchame, Malbet \& Monin 2003), or has assumed a gas
density and temperature distribution and solved for the chemistry 
(Willacy et al. 1998; Aikawa et al. 1999, 2002; Markwick et al. 2002;
Ilgner et al. 2004).
We solve for both the gas and dust temperatures and the chemistry simultaneously
in our models, and use the gas temperature to calculate the vertical
density profile. A similar 
theoretical study has been made by Kamp \& van Zadelhoff (2001), where the authors
attempted to explain the observed \mh emission from disks around two A stars, Vega and
  $\beta$ Pictoris as
seen by ISO and the lack of CO emission from these disks. These authors calculate the
gas temperature by balancing the different heating and cooling mechanisms, but
 do not solve for the vertical structure of the disk. Our work adds the computation of
vertical structure and is complementary to
their model, since we focus on the {\em inner} regions of the disk, within about
20 AU from the star where the gas and dust are both warm ($\sim$100 K),
whereas Kamp \& van Zadelhoff (2001) considered disk regions from \s $40-200$ AU. The inner
regions are more interesting  from the planet formation point of view, as many  planets are
expected to form close to the central star, within a few tens of AU if our solar system
is typical. Furthermore, gas in regions far beyond this radius may be subject to photoevaporation
from the central star and may be lost from the system early in its evolution. 
Our models are also complementary to the work of Glassgold,
Najita  \& Igea (1997) and Glassgold, Najita \& Igea (2004), who model the very
innermost regions  ($r < 1$ AU) of young optically thick disks  with
emphasis on the problem of heating the surface layers and the near-infrared spectra.

The paper is organized as follows.
In the next section, we
briefly describe some of the physical processes we consider. In \S 3,
we describe our model in detail. We present the results of a fiducial case (\S 4) and then
in  \S 5 present results where we vary most of the parameters to see how 
disk emission is affected by their values. This is followed by a discussion of
our results and conclusions (\S 6).
    
\section{Physical Processes in Disks} 
Our disk models include chemistry and thermal balance in a self-consistent
manner and calculate the density and temperature structure of the disk in 
both the radial and vertical directions. The vertical structure of the disk 
is calculated  by balancing the hydrostatic thermal 
pressure gradient with the vertical component of gravity from the central 
star. The inner ($r_i$) and outer ($r_o$) boundaries of the disk 
and the distribution of the gas and dust surface density $\Sigma(r)$  within these 
boundaries are input parameters. Our focus is on the IR emission 
detectable by the Spitzer Space Telescope, SOFIA or Herschel, and we therefore constrain our 
models to radial distances  \s $0.3-20$ AU from the central star, where gas 
temperatures range from \s 1000 K to 100 K respectively. 
The dust grains in this region are sufficiently warm to prevent ice formation 
on their surfaces, and therefore significant depletion of
relatively volatile species such as  gas phase O, C or 
S does not occur. We assume no significant opacity inside the inner edge
at $r_i$, so that there is an ``inner hole''. 

Although we focus on intermediate-aged and older debris disks which are relatively
 optically thin in dust opacity,
the gas opacity can in fact be quite large at line center of 
strong transitions and even for some FUV continuum bands, which lead to 
photodissociation and photoionization of abundant gas species. It is 
instructive to calculate a fiducial radial (hydrogen) gas column density $N_H$ in the 
equatorial plane. Assuming a gas surface density distribution $\Sigma(r) 
\propto r^{-3/2}$, a vertical scale height of $H=0.1r$, a gas mass 
of  \s 0.1 M$_J$, and an outer radius
$r_o = 20$ AU for a vertically isothermal disk, we find
\begin{equation}
N_H \approx 2.6 \times 10^{25} \left({{1AU}\over{r_i}}\right)^{3/2} 
\left({{M_{gas}}\over{0.1 M_J}}\right) {\rm\ cm}^{-2}. 
\label{col-est}
\end{equation}
Since cross sections in the UV and X-ray for trace species can approach $10^{-17}$ 
cm$^{-2}$ (or $\lesssim 10^{-21}$ per H nucleus), we see that even small masses 
of gas  can become optically thick in relevant
bands. We also note that most of the opacity for radially declining surface
densities is produced at $r_i$. If stellar photons make it through the
shielding at $r_i$, they will generally penetrate the rest of the disk.

We emphasize that our disk model is hydrostatic and our thermal and chemical solutions are 
steady state. Therefore, we implicitly assume that the timescales to reach chemical/thermal
steady state are short compared to dynamic timescales, i.e., the
timescale for radial or vertical migration of gas. Because of the high densities
and subsequent rapid cooling timescales, the assumption of thermal balance is
well justified. Chemical timescales, however, can be quite long in cool (T $\lesssim$ 300 K)
gas shielded
from photodissociation. In such gas, cosmic ray ionization sets the equilibrium timescales to \s
$10^6$ years, which is comparable to radial drift timescales. However, this work represents
an important first step in predicting the chemical and thermal structure of the warm
 T \s $100-1000$ K, $r \sim 1-20$ AU regions of evolved disks and predicting their spectrum. 

We briefly describe below the chemical and physical processes considered in our
models and refer the reader to the appendices for greater detail. 

\subsection{Dust Physics}
Coagulation processes in intermediate-aged disks may lead to a distribution of solid 
particles which range in size from 1 $\mu$m to 10$^{4}$ km (planets). 
We define ``dust'' to consist of all particles $\le a_{max}=1$mm in
radius. We assume that the size distribution of particles between $a_{min}$ and
$a_{max}$
follows a power law $n(a)\propto a^{-s}$, where $3<s<4$ such that most of 
the surface area is in the smallest particles and most of the mass is in 
the largest particles. Observational studies of disks estimate the 
``dust'' mass from emission at wavelengths $\simeq 0.4-2 $ mm. If the size
distribution has the above power-law form, these observational studies 
are only sensitive to the mass of dust in particles with sizes $\lesssim$ 1 mm. 

Implicit in these assumptions is that most of the mass of solids
will be ``hidden'' in large particles, rocks, boulders and planetesimals
which emit insignificant amounts of infrared and millimeter wavelength
radiation. Thus, at intermediate disk ages ($\sim 10^7$ years), where
dust has coagulated up to planetary sizes but where significant quantities
of gas may not yet have been 
dispersed, the gas to ``dust'' (meaning $a\lesssim 1$mm) mass ratio may increase 
above the value \s 100 characteristic of interstellar conditions. 
For example, if 90\% of the mass of the primordial dust in a disk 
grows by runaway accretion to cores/embryos of sizes $10^3$ to $10^4$ km, and the remaining 10\%
of the mass is in  solids with a distribution characterized by $a_{min}=1\mu$m,
$a_{max}=100$km, and $s=3.5$, the gas to ``dust'' mass ratio increases
to $10^7$ without any gas or solid dispersion.
Later, as the gas disperses, the gas to dust mass ratio may fall below the
interstellar value.
 
In order to express the large range of gas to (small) dust mass ratios and the
absolute values of the gas or dust mass which may exist as intermediate-aged
or older  disks evolve, we
consider a range of gas masses $M_{gas} \simeq 10^{-3} - 1$ \mj\ and a range of
dust masses $M_{dust} \simeq 10^{-7} - 10^{-4}$ \mj\ (or $f_D \sim 10^{-4}-10^{-1}$). 
$M_{gas}$ and $M_{dust}$ are treated as independent 
variables, so that we cover gas-to-dust mass ratios of $10^7$ to $10$. We
emphasize that these masses lie at $r\lesssim$ 20 AU. We do not
consider gas masses smaller than $10^{-3}$ \mj\ because we find such small
masses to be undetectable by the Spitzer Space Telescope, SOFIA or Herschel. We do not consider dust
masses higher than $10^{-4}$ \mj\ because we find that the dust  becomes 
significantly optically thick in the equatorial plane, and our model is
valid only for small to moderate dust optical depths.

Our models consider gas and dust separately but as a coupled system, and they 
are capable of treating the dust grain density distribution independently of the 
gas density distribution as a function of radius $r$ and vertical height $z$. In 
addition, the grain size distribution can be arbitrarily varied with $r$ and $z$. 
However, for simplicity and to limit the number of models to be
presented, we make the following assumptions.
\begin{enumerate}
\item We assume that the dust mass density distribution $\rho_{d}(r,z)$ is 
proportional to the gas mass density distribution $\rho(r,z)$. That is, for 
a given disk we assume that the same gas-to-dust mass ratio applies everywhere 
in the disk. This assumption means that we ignore settling by grains to 
the midplane. Turbulence may stir the gas and dust
sufficiently to justify this assumption (e.g. Balbus \& Hawley 1991). However,
our main motivation is simplicity in the models; presumably by using an average (vertical)
mass ratio of gas to dust, we will obtain a first order solution for the more 
complicated case of vertical differentiation.
\item We assume that the  grain size distribution does not vary with $r$ or $z$. 
We consider our minimum grain size to be much
larger than that in the interstellar medium.
In intermediate-aged disks, the population of very small grains
dwindles as coagulation processes lead to grain growth (e.g., 
Beckwith \& Sargent 1996; Throop et al. 2001; Natta et al. 2004).  
In more evolved debris disks,  
small grains are continually replenished  due to grinding 
down of larger objects by collisions, but radiation pressure forces act on
sub-micron sized dust grains to rapidly remove them from the
disk (Backman \& Paresce 1993) on short timescales. Observations
nevertheless seem to indicate the presence of small sub-micron
sized particles (Li and Greenberg 1998) in a debris disk. We
adopt a minimum grain size of $a_{min} = 1 \mu$m for our
standard case, but also consider other cases where we vary
$a_{min}$. 
In this paper, we only consider $n(a) \propto a^{-3.5}$, i.e.,
$s=3.5$.
\end{enumerate}

For a given mass 
in grains, the surface area $\propto a_{min}^{-1/2}$. Gas-grain collisional
heating can be important in disks and
due to their large surface area, small grains dominate the heating of the gas. 
Therefore, lowering $a_{min}$ raises the gas-grain coupling and usually 
raises the gas temperature in the disk.
Lowering $a_{min}$ tends to raise the gas temperature for two other
reasons. 
Smaller grains  have higher efficiencies of heating the gas through the grain 
photoelectric heating mechanism (see Watson
1972; Weingartner \& Draine 2001), which may dominate the heating
once $a_{min} \le 10^{-2} \mu$m. More importantly for this paper, 
smaller grains are hotter than larger grains due to the heating by 
stellar radiation and the inefficient radiative cooling of small grains. 
Grain absorption and emission efficiencies are dependent primarily
on their physical size relative to the wavelength of emission or 
absorption, and their composition. For our models, we choose silicate
(amorphous olivine) grains and the absorption coefficients are 
calculated from a Mie code (S.Wolf, private communication) 
based on optical constants from the Jena database (Henning et al. 1999). 
We  divide the grain size distribution
into 32 logarithmically spaced bins in size $a$, and each size grain 
has a different temperature at a fixed point in space. 

We conclude this subsection with a discussion of the dust optical depths
in the model disks.  Using the
absorption coefficients discussed above, we calculate the optical depth $\tau_V$ in the 
V band (0.5 $\mu$m) of dust grains radially outward in the equatorial plane of the disk.
For most of our models,  $\tau_V \lesssim 1$. We do, however, run a few
cases where the optical depth in the midplane $\tau_V \sim 5$. These disks
are still optically ``thin'' in comparison to the younger, primordial dust
disks where $\tau_V \sim 10^3-10^4$. In our calculation
of dust temperatures we ignore the heating of dust grains due to the infrared
radiation field emitted by the surrounding dust, but
we include extinction of
starlight by dust grains along the line of sight to a particular point in the
disk. Even at optical depths $\tau_V=5$, the dust grains are still primarily
heated by attenuated starlight rather than by the re-radiated IR continuum field,
so that our error in calculating the dust temperature is still small ($\lesssim 10\%$),
and our models are still valid. Typically, debris disks ($t>10^7$ years) have
$\tau_V \lesssim 10^{-2}$. However, intermediate-aged disks ($t \sim 10^{6.5}$ years)
may have $10^{-2} < \tau_V <5$.

\subsection{Gas Chemistry}
The chemical composition of the gas disk is critical to the structure and
evolution of the disk. Although \mh cooling is often important in the radial 
range we consider, trace species such as S, O, CO and H$_2$O are often
significant or even dominant coolants
and for an accurate determination of the strength of emission lines, 
the abundances of atoms and molecules need to be calculated.
 Our treatment of disk chemistry is an adaptation of the Photodissociation
Region (PDR) models of Tielens and Hollenbach (1985, hereafter TH85) and 
subsequent work (Kaufman et al. 1999). Elemental gas-phase abundances 
are taken from Savage \& Sembach (1996) (Table~\ref{abun}, Appendix A).
We use interstellar depletions of these elements, allowing for
refractory dust but no ices in the $r\le20$AU disk region, where the dust is
sufficiently warm to prevent ice formation.
We have added sulfur chemistry 
and reactions of CH$_3$ and CH$_4$, and there are a total of 73 species 
with atoms and ions of H, He, C, O, S, Fe, Mg, Si and molecules and molecular
ions of H, C, O, S and Si (Table~\ref{species}, Appendix A). 
To the PDR network of ion-neutral, neutral-neutral,
and photoionization reactions, we have added X-ray photoreactions,
and some high temperature reactions of importance under typical disk conditions.
We  have 537 reactions in all, with reaction rates in most cases 
taken from the UMIST Astrochemistry database (Le Teuff et al. 2000). A 
list of the added reactions, and the rates when they differ from the
UMIST data, can be found in Table~\ref{rates} (Appendix A). 

The formation mechanism of \mh in disks may differ from the processes
at work in the interstellar medium. In the ISM, \mh forms mainly on the
surfaces of cold dust grains, with $T_{dust} \lesssim $  25 K, typically. Dust in
disks at $0.3-20$ AU typically have temperatures $T_{dust} \gtrsim 100 $ K.
It is not clear if H atoms  would stick to such 
warm dust grains for sufficient time to allow formation of
\mh molecules (Hollenbach \& Salpeter 1971; Cazaux \& Tielens 2002). 
We have conservatively omitted grain formation of \mha. We find that 
other processes such as three-body reactions in dense 
regions and formation through H$^{-}$ can completely
convert atomic hydrogen to molecular hydrogen in self-shielded regions. 
We have also run cases including grain formation of \mh and find that the
resultant infrared spectrum is not significantly altered. 

Because the dust is optically thin, stellar and interstellar UV radiation is 
mainly attenuated in the disk by  different gas species in various bands and lines.  
Large gas columns in the inner disk shield the outer regions from 
UV photodissociation caused by the central star. Similarly, the gas columns
in the upper layers shield the midplane from the interstellar radiation
field (ISRF). 
The self-shielding of \mh and CO, which involves absorption of specific
resonance lines rather than continuum radiation, is treated analytically as in
TH85. Attenuation of continuum FUV radiation by atoms and molecules
is determined by splitting the FUV
stellar radiation field into nine energy intervals, and the photon flux
in each interval is calculated from the stellar (and ISRF) spectrum. The nine intervals 
 from 0.74eV to 13.6eV include 0.74$-$2.6eV, 2.6$-$3.5eV, 3.5$-$4.3eV, 4.3$-$5.12eV, 5.12$-$7.5eV,
7.5$-$10.36eV, 10.36$-$11.26eV and 11.26$-$13.6eV where the lower and
upper bounds in each case were chosen to correspond to dominant
photoabsorption thresholds of species in each energy bin.
The attenuated photon flux in each bin  at a given position 
in the disk is dependent on the column densities of all the absorbing species 
towards the line-of-sight to the source.  Photodissociation and photoionization 
reaction rates are thus accurately computed by taking into account the
attenuation of photons in the energy range where absorption cross-sections peak.

We also consider the effects of X-ray radiation from the star on the disk.
X-ray luminosities of \s $10^{-4}$ L$_{bol}$ have been observed in stars
as old as $10^7-10^8$ years, and therefore include not only intermediate-aged disks
but also a number of observed ``debris disks''.
X-rays affect the chemical structure of the disk through ionizations
and heat the gas via collisions with secondary electrons generated
by the  X-ray produced primary electron. 
X-ray heating and chemistry may dominate in the inner regions of the 
disk and the surface layers where attenuation by gas is low.
Stellar X-ray spectra from weak-line T Tauri stars 
peak at about 1$-$2 keV (e.g., Feigelson and Montmerle
1999) and the flux drops rapidly as the X-ray photon energy increases.
 Soft X-rays,
with energies $\lesssim$ 1 keV, do not penetrate significantly through a 
massive disk, their typical attenuation columns being about 
$10^{21-22}$ cm$^{-2}$ of gas.
The penetration columns are larger by about two orders of magnitude for
harder (3$-$10 keV) X-rays, but the stellar flux and photoabsorption 
cross-sections  decrease sharply to make
their influence on disk chemistry relatively unimportant in the 
shielded regions of the disk. On the other hand,
in disks with low masses or lower column densities towards the line
of sight to the star (disks with large inner radii, for example),   
X-rays incident on the disk heat, ionize and dissociate significant numbers
of atoms and molecules in the
gas and our chemical code treats this chemistry in detail. 
For simplicity, we do not consider double or higher ionizations, 
as in Maloney et al. (1996).   The total X-ray photoabsorption cross section
of the gas is taken from Wilms, Allen \& McCray (2001), adapted to our
gas-phase abundances (see Appendix B).
We use the photoionization cross sections as originally published by
Verner et al. (1993) and later updated by Verner \& Yakovlev (1995).
The primary  ionization rates for each species are calculated for each
species by integrating the product of the attenuated photon flux and the absorption
cross section over the X-ray energy range (Maloney et al. 1996). We also
include secondary ionization rates and  X-ray induced FUV radiation and 
the resulting photochemistry 
(Maloney et al. 1996). The details of our treatment of X-ray chemistry 
and heating are given in Appendix B.

\subsection{Heating and Cooling Processes} 
One of the major innovations of this paper in terms of disk modeling
is the separate calculation of gas and dust temperatures. Most previous
models (with the exception of Kamp \& van Zadelhoff 2001) assume that the
gas temperature equals the dust temperature. While this is certainly
true in the midplanes or central regions of very optically thick disks, it is
not as true for debris disks, intermediate-aged disks, 
or the upper regions of younger, optically
thick disks. Here, the relatively weak thermal coupling between gas and dust, and the
different processes which heat and cool the gas and dust, 
can lead to significant temperature differences.  This point was
also made by Chiang \& Goldreich (1997), who note that the gas temperature
in the superheated layers of passive, optically thick dust disks 
can be significantly lower than the dust temperature due to molecular cooling. 

Gas in disks can be heated through many different mechanisms,
and we consider the physical and chemical processes that are likely to
dominate in these environments. The heating of dust grains by the stellar radiation
field and subsequent collisions with gas molecules transfers kinetic energy to the
gas thereby heating it. Gas-dust collisions are the dominant heating mechanism  
for the gas through most of the disk, except in regions where the gas is heated
to temperatures higher than the dust by other processes and  here collisions
with dust  cool the gas. In regions where the columns through the disk are
small enough to allow the penetration of FUV radiation, and \mh is photodissociated
to form atomic hydrogen, the heat associated with the formation of \mh
 raises the temperature of the gas. X-rays
from the central star directly heat the gas and dominate the gas heating at the surface
and in the inner disk. In addition to these mechanisms, we also  include
 heating due to collisional de-excitation of vibrationally-excited 
\mh molecules, grain photoelectric heating, drift heating due to dust, exothermic chemical 
processes, photoionization of neutral carbon, 
and cosmic rays. All the heating mechanisms considered are described in 
some detail in Appendix C. 
  
The gas in the disk is mainly cooled through radiative transitions
of  the different species and, where the gas is warmer than the
dust, through gas-grain collisions.
We include cooling due to the fine structure and  metastable lines of
CI, CII, OI, OII, FeI, FeII, SI, SII, SiI, SiII, the
rotational lines of H$_2$, H$_2$O, OH, CO, CH, HCO$^{+}$, and by Lyman $\alpha$. 
We also include cooling due to the collisional ionization of atomic species,
and vibrational cooling due to \mha. Vibrational cooling  due to CO
was found to be unimportant for the disk parameters considered in this paper.
For all the molecular rotational
levels, except for H$_2$ and CO, we use the analytical formulae of 
Hollenbach \& McKee (1979, hereafter HM79). We derive new fits to
their formulae for H$_2$O (ortho and para states) and OH (see Appendix D)
using collisional rate data for 167 and 45 levels respectively. It was 
computationally expensive to include a detailed multi-level 
calculation for these molecules in the code, and we estimate the 
strengths of individual H$_2$O and OH lines using the derived density and
temperature structure and solve for the level populations after a disk
solution for the temperature, density and chemical abundance structure
(i.e., the $r,z$ dependence) has been obtained with the new analytical 
approximations for the total cooling by OH and H$_2$O molecules. We note that
cooling transitions are often optically thick, and we use the escape
probability formalism described in HM79.

We solve for the rotational cooling due to \mh in detail because
one of our primary aims is to calculate the strength of the \mh
emission lines from the disk. We include the first 20 levels of molecular hydrogen 
in our calculation of the cooling, and derive the population of each 
individually excited level. We have also added vibrational cooling of \mh
as in HM79 and cooling due to collisional dissociation of \mh molecules. We
do a similar calculation of the first 20 rotational levels of CO and solve
for the level populations to compute the total CO cooling. 

\section{Disk Model}
In this section, we describe our disk model, discuss some
of our assumptions and give the details of our numerical scheme
for obtaining a disk solution. Because our main aim is to characterize observed
emission from disks and infer disk properties from the comparison of
observed spectra with our models,
we have varied most of our input parameters
 across a wide range. However, we define a ``standard
case'' where we assign typical values to each parameter.
The results from the standard model are  
described in \S 4.  
\paragraph{Surface density distribution}
We assume a power law distribution of surface density of both gas and
dust with radius,
$\Sigma(r) \propto r^{-\alpha}$, where $r$ is the distance from the
star. The nominal value for $\alpha$ is often taken to be $3/2$
in the literature (Lynden-Bell \& Pringle 1974; 
Chiang \& Goldreich 1997). For our standard case, we assume a flatter
surface density distribution defined by $\alpha=1$, motivated
by observations of disks (Dutrey et al. 1996, Li \& Greenberg 1998). 
The main effect of changing the slope
of the surface density distribution is to alter the relative distribution
of mass in the inner and outer disk. This changes both the slope
of the dust continuum emission at longer wavelengths and the relative strengths
of the gas emission lines that  preferentially originate either in the
warmer inner disk or the colder outer disk region. We study the effects of
a change in the surface density distribution by varying $\alpha$.
As long as $\alpha<2$, most of the disk mass is in the outer regions;
and for $\alpha>0$ (and assuming a vertical scale height $H$ that increases
with $r$),
most of the radial opacity occurs in the inner regions.
\paragraph{Radial extent}
Our model disk is defined to have both an inner and outer edge.
It is not, however, certain whether an inner (gas) hole would exist 
in evolved  disks, even if the dust is heavily depleted in the
inner regions.
Theoretical models of dust disks often assume 
an inner truncation radius determined by the dust sublimation 
temperature, interior to which there is no dust (e.g., Bell 
et al. 1997; Chiang \& Goldreich 1997; D'Alessio et al. 1998;
Dullemond et al. 2002). SED modeling of observed dust disk emission 
also argues for the presence of inner (dust) holes in many
disk systems (e.g., Natta et al. 2001, Dullemond et al. 2001).
In evolved disks, where planetary formation is likely to have
taken place, an inner gas cavity may be generated by massive planets
that prevent outer gas from accreting onto the star, whereas
viscous accretion of gas interior to the massive planets
may rapidly drain gas there.

 The inner disk 
radius $r_i$ is an important parameter of our models because it
determines the penetration of stellar radiation into the disk
and  contains the hottest gas and dust. Emission from this region, therefore,
is a dominant contributor to the higher frequency continuum from the 
dust and the higher excitation lines of the gas. Small values
of $r_i$ imply higher densities and column densities
 at the inner edge and radiation can
penetrate only a short distance into the disk before being
attenuated. If disks with the same gas mass have larger $r_i$, there is less
rapid attenuation of stellar photons.  Consequently, photodissociation
and photoionization by X-rays and FUV radiation extends 
further into  the disk and affects more of the gas mass. 
Observations of intermediate-aged and debris disks indicate
disk inner radii ranging from a few tenths of an AU to 
tens of AUs (e.g., Liu et al. 2004, Calvet et al. 2002,
Li \& Greenberg 1998). 
For our standard case we assume an inner radius of $r_i=1$ AU. 
However, we include  several cases where we vary the inner radius
to see how this parameter affects disk thermal structure and chemistry
and the resultant spectrum.

We also define an outer disk radius $r_o$. Our fiducial disk
is truncated at $r_o = 20 $ AU. Unlike $r_i$, the exact choice of $r_o$
does not affect our results significantly because any gas beyond
$r_o$ will be too cold to significantly contribute to the IR lines we focus
on in this study. The choice of 20 AU is motivated by the
following four reasons. $(i)$ Disks may lose much of the outer gas through
the process of photoevaporation early in their evolution
(Hollenbach et al. 1994; Adams et al. 2004). $(ii)$ Our interest lies in the 
planet-forming regions of the disk, and from our understanding of
the formation of the solar system and of the evolution of disks,
 planet formation appears to take place 
within a few tens of AU from a star. $(iii)$ Glassgold et al. (2004) present
models of inner ($<$ 1 AU) disks whereas Kamp \& van Zadelhoff (2001) present
models of outer ($\ge 30$ AU) disks. We, therefore, present this complementary
study, which is especially relevant to the Spitzer Space Telescope, SOFIA and Herschel since the gas
and dust temperatures span a range ($\sim$ 50 K - 500 K) which produce 
copious IR emission in the 4 $\mu$m $-$ 200 $\mu$m wavelength range.
Mid-infrared emission from gas species at temperatures less than
$\sim 100$ K is likely to be insignificant, so the neglect of any
gas beyond $r_o$ does not affect our emission line calculations. 
$(iv)$  Gas molecules, such as H$_2$O,
may freeze on grain surfaces when the dust grain temperature drops 
below $\sim 80-100$ K.  We therefore truncate the disk at a radius
where  ice mantle formation is likely to become important on grains, i.e., when
dust grain temperatures fall below $\sim$ 100 K. 
To summarize points $(iii)$ and $(iv)$, the outer edge need not correspond
to a real cut-off in surface density. We are interested in the IR emission
from warm gas in the $r\lesssim 20$ AU region. Material beyond this outer
radius will be either too cold or frozen on grain surfaces to radiate
appreciably as IR line emission from gaseous species. 

\begin{figure}[f]
\plotone{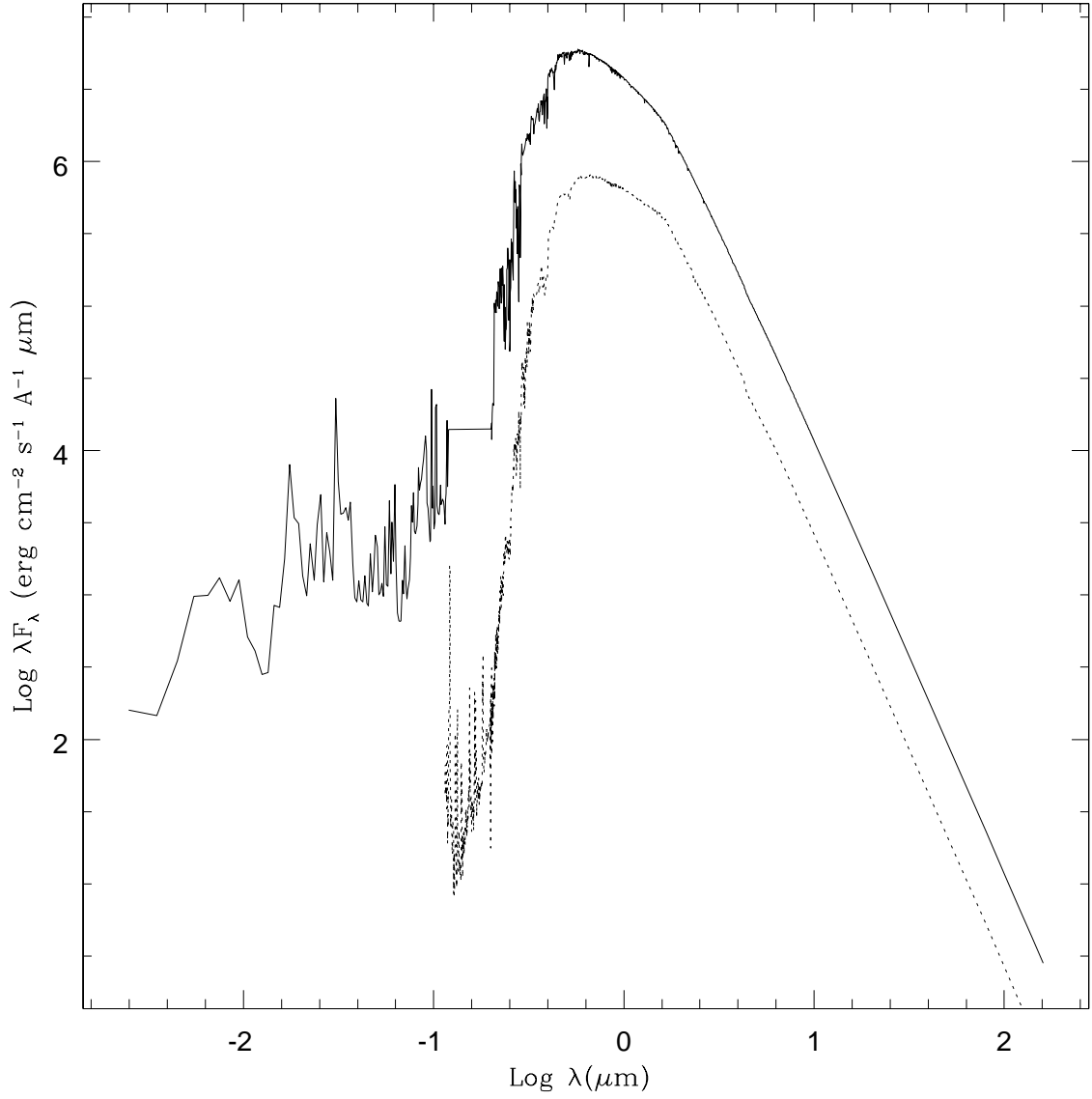}
\caption{The adopted stellar spectrum for a 10 Myr old G2  star, consisting
of an age-dependent UV flux added to a Kurucz model for the same spectral-type.
The dashed line is a spectrum for a K star that includes chromospheric emission.} 
\label{stellar-flux}
\end{figure}

\paragraph{Stellar Spectrum}
The stellar spectrum is chosen to be a modified Kurucz model, and
we consider stars of two spectral types, a G star at 6000 K and a 
cooler K star with a temperature of 4000 K.  
The central star at these disk ages is
expected to have a significantly higher FUV luminosity than its
main-sequence counterpart  and we account for this increased
luminosity in our stellar models. An accurate determination of the
FUV flux is especially important in solving for the 
disk chemistry as the FUV initiates numerous photodissociation and 
photoionization reactions. We follow a prescription by Kamp \& Sammar (2003)
for our age-dependent FUV fluxes, where the FUV flux scales 
inversely with the age of the star (Ayres 1997). We assume an age of
$10^7$ years for the G star in our models. IUE and rocket experiment
data are added to a Kurucz photosphere model to thus
obtain a semi-empirical age-dependent spectrum for a G2 spectral type star. 
We use chromospheric models by Cohen(private communication) for the star 
$\epsilon$ Eri as our input stellar model for the K star. Figure~\ref{stellar-flux} shows 
the spectra we use, the solid line indicates the G star and the dashed line the
K star. 

For the stellar X-ray spectrum, we use a broken power-law fit to
ROSAT observations (Feigelson \& Montmerle 1999)
for a wTTS, where the luminosity per keV peaks at about 2 keV. The
luminosity per keV is thus given by $L(E) \propto E^{-1.75},
E \ge 2$ keV and $L(E) \propto E$ for $ E < 2$ keV. This X-ray spectrum 
is normalized to the stellar X-ray luminosity between 0.5$-$10 keV.
For our fiducial case, the X-ray luminosity is assumed to be
$10^{-4}$ times the bolometric luminosity of the star. 

\paragraph{Numerical model}
The disk density and temperature structure is calculated iteratively
from the midplane to the surface starting at $r=r_i$ and then
working radially outward in the disk.
We start with an estimate of the  gas density at the point ($r=r_i,z=0$) and
compute the dust and gas temperatures.

The dust temperature is calculated from thermal balance between the 
absorption of stellar radiation by the dust and subsequent re-emission in the
infrared. Heating of a grain due to absorption of infrared radiation 
coming from surrounding dust particles is ignored in our optically thin
dust disk models.  The dust temperature (at a given size $a$)  needs to
be iterated only because the dust vertical distribution is determined by the 
gas vertical distribution, and the dust temperature is a function of 
$z$ because of the varying distance to the star. 

The gas temperature
is determined by thermal balance between the heating and cooling
processes described earlier, and this then determines the gas pressure.
The condition of vertical hydrostatic pressure equilibrium,
\begin{equation}
{{dP}\over{dz}} = {{-G M_* \mu m_H n_H(r,z)}\over{(r^2+z^2)^{3/2}}}
\label{presbal}
\end{equation}
determines the pressure as a function of disk height, where $P=
k_B (n($H$)+n($H$_2)+n($He$)) T_{gas}$, $\mu$ is the mean molecular weight,
$m_H$ is the mass of a hydrogen atom,
 $n_H$ is the hydrogen nucleus number density, and
$n($H$)$, $n($H$_2)$ and $n($He$)$ are the number densities of atomic and molecular 
hydrogen and helium respectively. 
At each height $z$ from the midplane, the density
and temperature combination that gives the pressure from Eq.(\ref{presbal}) is
found. The vertical
grid is adaptive in nature and  determined by the condition of pressure 
gradient balancing gravity. The vertical increments in height are chosen to correspond 
to equal factors of decrease in  thermal pressure (a factor of 0.8
in most cases) and for regions where the temperature slowly varies with 
$z$, this also corresponds to a logarithmically decreasing column 
density in each cell. This ensures that the escape probabilities for the
cooling transitions increase slowly and smoothly
as we get farther from the midplane, and enables accurate determination of
the gas temperature. We ensure that the density contrast from surface to 
midplane is at least four orders of magnitude, and the number of cells and
their spacing are calculated to meet all these criteria. 
 After the entire vertical grid has been solved, the gas
density is integrated over $z$ and compared to the
required gas surface density, $\Sigma(r)$. 
The midplane density is re-scaled and the vertical temperature
and density structure is thus iterated  until
the surface density converges to the required value.

Once we have the solution at $r_i$, we move outward to solve at the
next radial grid point.
The radial grid is logarithmically divided in $r$, except for the 
inner disk which is more finely sub-divided. The column density
towards the star is dominated by the dense inner regions, and 
this is where the gas absorbs most of the stellar FUV and X-rays.
This attenuation needs to be calculated accurately, and
we have divided the inner disk into cells of slowly increasing column
density going outwards in $r$, with the first radial cell (at midplane)
having a hydrogen gas column density $10^{20}$ cm$^{-2}$, which
corresponds to a small attenuation of the X-ray and UV fluxes. 

\section{Standard Disk Model Results}
In this section, we  describe in some detail the results of our 
fiducial disk model (``the standard case''),
the input parameters for which are listed in Table~\ref{standard-par}.
\begin{table}[f]
\caption{Input Parameters for Standard Disk Model}
\label{standard-par}
\begin{tabular}{l r l r}
\hline 
Gas mass & $10^{-2}$ M$_J$ & Dust mass & $10^{-5}$ M$_J$ \\
Inner radius $r_i$ & 1 AU  & Minimum grain size $a_{min}$ & 1.0 $\mu$m \\
Outer radius $r_o$ & 20 AU   & Maximum grain size $a_{max}$ & 1.0 mm\\
Surface density power law index $\alpha$ $$ & 1.0 & Grain size dist. index $s$
& 3.5 \\
Stellar age (for FUV flux) & $10^7$ years & Grain composition & Am. Olivine \\
Stellar X ray Luminosity & 4.0 10$^{29}$ erg s$^{-1}$ & Stellar bol. luminosity
&  4.0 10$^{33}$ erg s$^{-1}$ \\
Stellar temperature $T_*$ & 5780 K &&\\ 
\hline
\end{tabular} 
\end{table}
The standard model has a gas mass $M_{gas}=10^{-2}$ M$_J$ and dust
mass $10^{-5}$ M$_J$. 
In the next section we present  a full parameter study varying both the gas
mass and the dust mass, the main parameters of this work. 
 The other parameters we treat as ``minor'' parameters,
and in the next section we vary each of these parameters in turn, holding all other parameters
at the fiducial values, and study its effect on disk structure and emission. 
These minor parameters include 
the disk inner radius, the minimum grain size, the stellar
X-ray flux and the surface density distribution.
 
Our fiducial gas to dust mass ratio of $10^{3}$
is higher than the interstellar value, but, as we have discussed, this
is possible at an intermediate stage where dust has coagulated but gas
has not yet dispersed. We have chosen this value for our fiducial case 
primarily because, as shall be shown below, this ratio allows detection of 
gas emission lines over the continuum for the smallest gas masses with the
dust midplane optical depth still being $\lesssim 1$. 

\begin{figure}[f]
\plotone{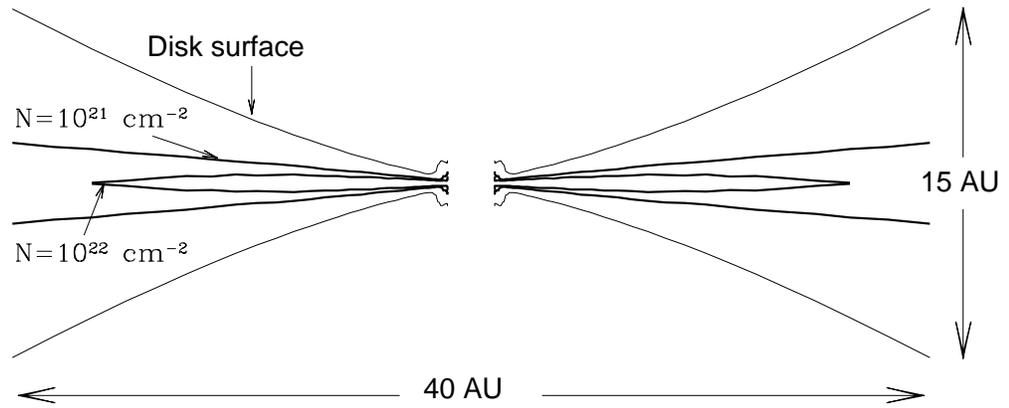}
\caption{Disk surface for the standard case  at a density equal to
$10^{-4}$ the density in the midplane, and contours that indicate
surfaces where the vertical gas column densities to the
surface are $10^{21}$ and $10^{22}$ cm$^{-2}$. }
\label{disk}
\end{figure}

Our standard model disk solution results in a flared disk with an inner puffed region where
the gas is subject to the unattenuated, strong stellar radiation field
and heated to high temperatures. Dullemond et al. (2002) find
a similar effect in younger, optically thick (in dust) disks. Figure~\ref{disk}
shows the disk surface where the density is $10^{-4}$ times the density in the
midplane and contours where the vertical column densities to the surface are
$10^{21}$ and $10^{22}$ cm$^{-2}$.  Stellar UV and X-ray fluxes are 
absorbed at these column densities, and hence in these regions.  
The hot inner edge or rim
has a somewhat lower density compared to gas just beyond the edge because 
it is hotter and more extended vertically.
This region shields much of the rest of the  disk from UV and X-ray photodissociation
and photoionization by starlight. 
The density increases slightly with radius progressing from this inner edge
as the temperature falls steeply, and the disk scale height decreases. 
Recall that for a vertically isothermal disk  the scale 
height $H$ of the disk is given by $H \propto  (r^3\ T(r))^{1/2}$ 
(e.g., Adams, Shu \& Lada 1988, Shakura \& Sunyaev 1973). Therefore
for a radial temperature gradient steeper than $1/r^3$, the scale height decreases
with radius. Further into the disk, the temperature falls less steeply and here
$H/r$ now increases with radius, giving rise to a flared disk
solution. The density falls less steeply with height as compared to an
vertically isothermal solution in most regions of the disk, because the gas temperature
in general rises  with $z$.  We first describe 
the main heating and cooling mechanisms in the disk and the 
resulting temperature structure and then discuss the  chemistry 
and  emission from the disk.

\subsection{Heating, Cooling and Temperature Structure} 
\begin{figure}[f]
\plotone{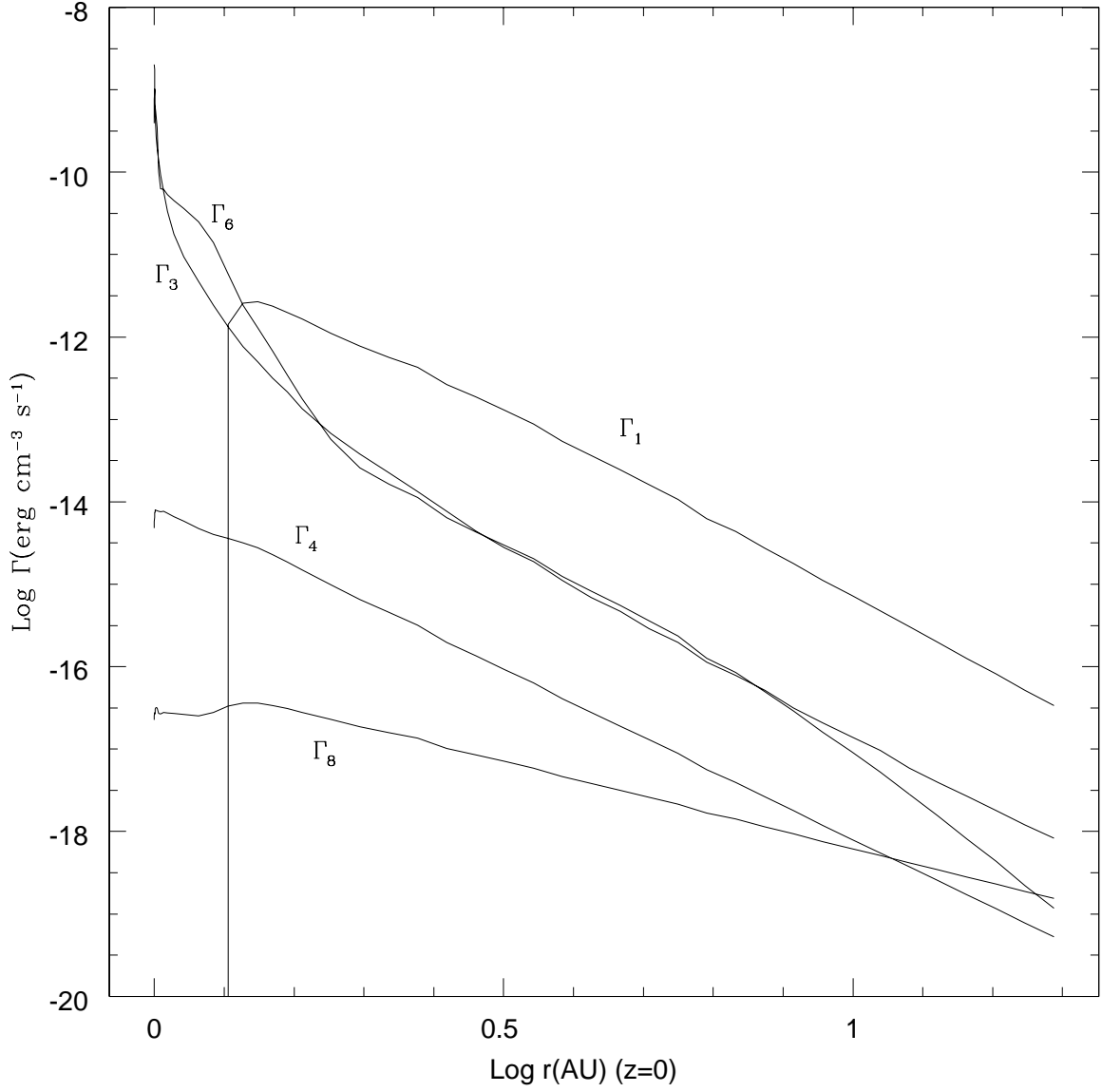}
\caption{Heating terms as a function of radius in the disk midplane for the
standard disk model with $M_{gas}=10^{-2}$ \mj\ and $M_{dust}=10^{-5}$ 
\mj. $\Gamma_1$ is the heating due to gas-grain collisions, $\Gamma_3$ is
X-ray heating, $\Gamma_4$ grain photoelectric heating, $\Gamma_6$ 
heating due to H$_2$ formation and $\Gamma_8$ cosmic ray heating.
See Appendix C for details.}
\label{heat}
\end{figure}

\begin{figure}[f]
\plotone{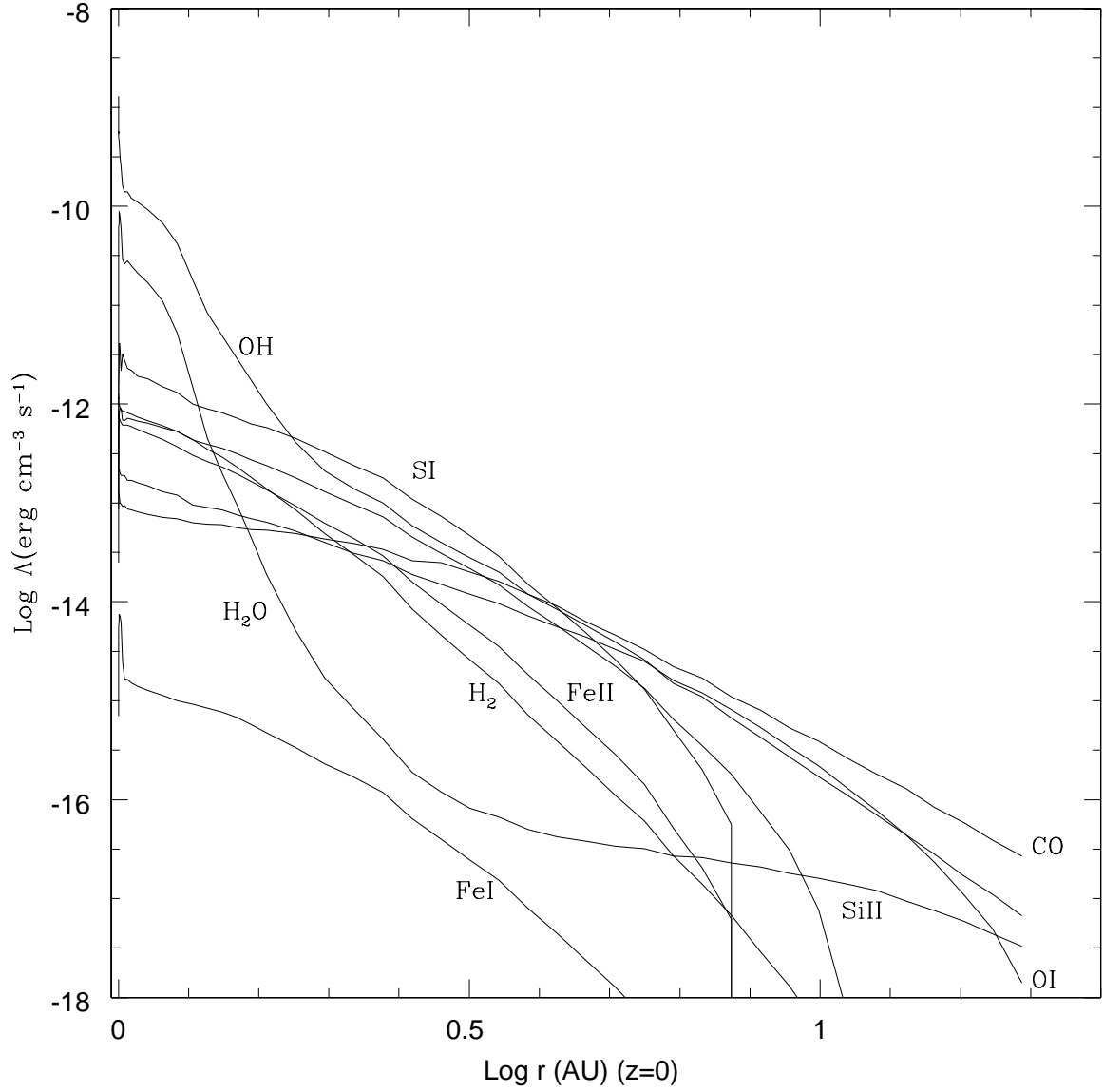}
\caption{Cooling due to different chemical species as a function of radius
in the disk midplane for the standard disk model.}
\label{cool}
\end{figure}

\begin{figure}[f]
\plotone{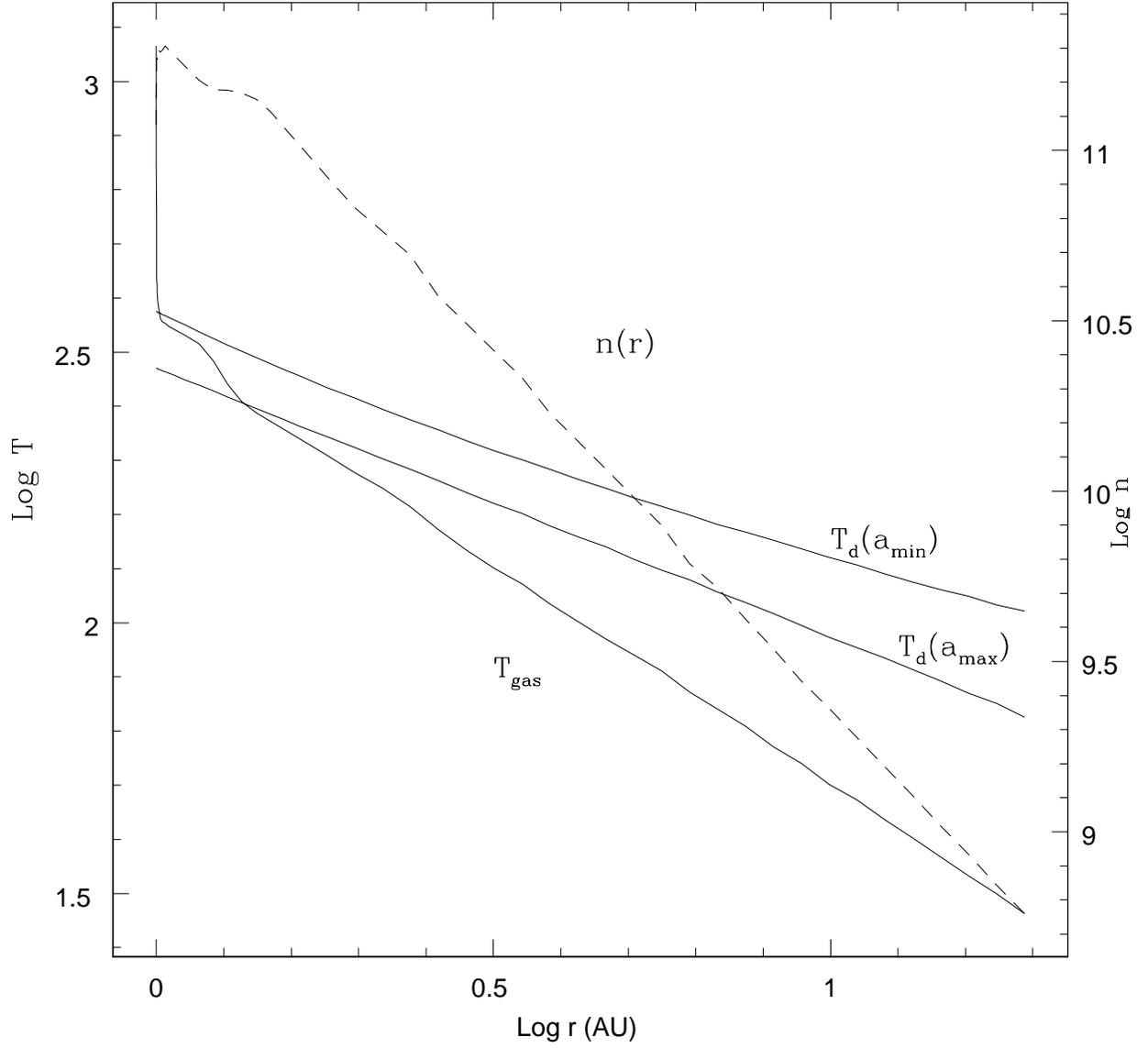}
\caption{Gas and dust temperatures in the midplane of the disk are shown as
(labelled) solid lines, whereas the dashed line shows the gas number density in the
midplane for the standard disk model.}
\label{tempr}
\end{figure}

\begin{figure}[f]
\plotone{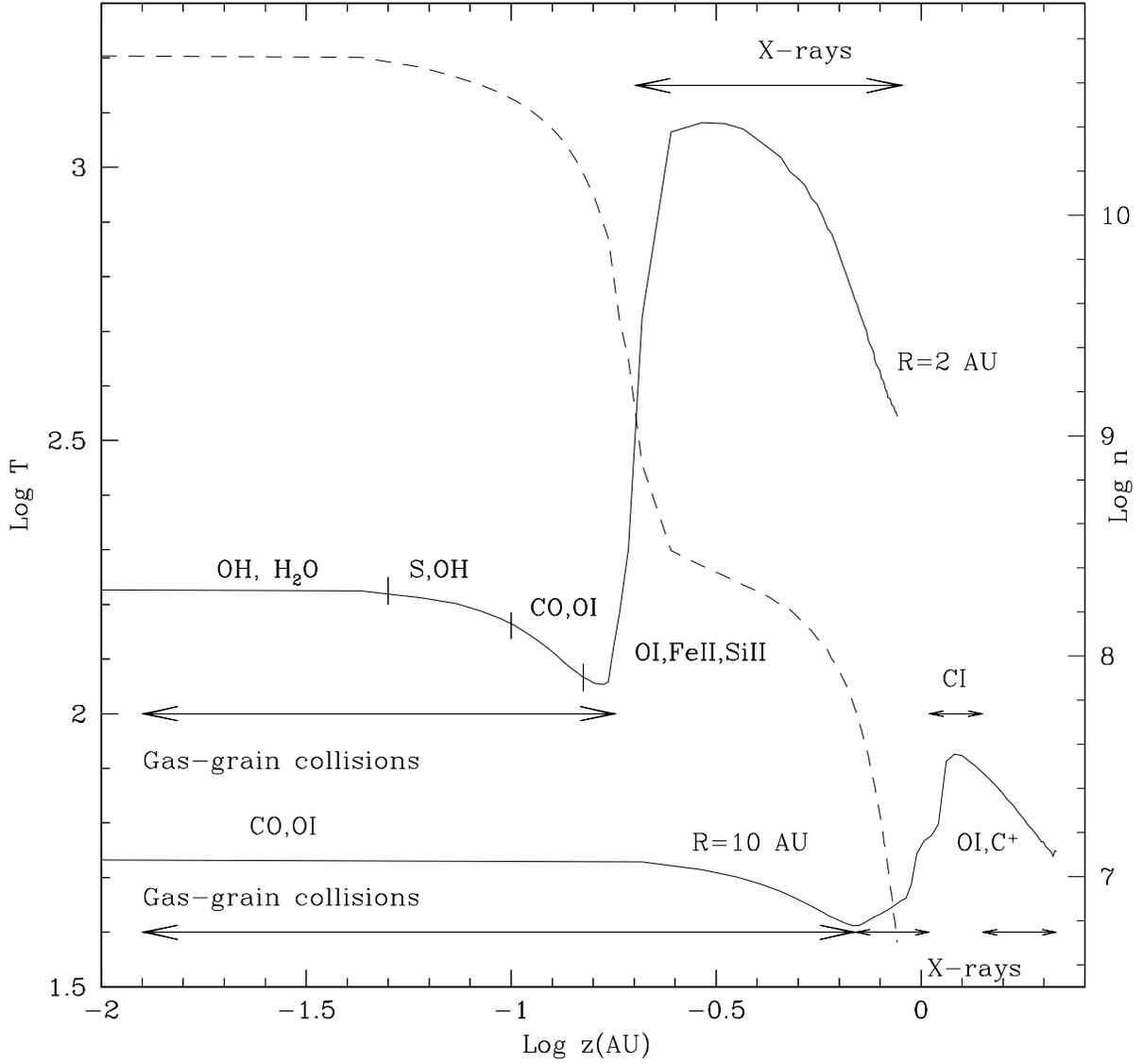}
\caption{Gas temperature (solid line) and density (dashed line) as a function of z
at a radial position of 2 AU and the temperature at 10 AU for the
standard disk model. The dominant heating
agents and the coolants in each region are marked on the temperature curve.}
\label{tempz}
\end{figure}

Figures~\ref{heat}, \ref{cool}, \ref{tempr} and \ref{tempz} show the
heating, cooling and the radial temperature and density structure
in the disk at the midplane and the vertical temperature and density
structure in the disk at two typical radii, of 2 and 10 AU. 
Figure~\ref{heat} shows the main heating terms for our standard case in the midplane of the
disk, as a function of disk radius. Heating due to formation of \mh is
high where hydrogen is predominantly atomic, but quickly decreases
as self-shielding turns the gas molecular, at hydrogen nuclei column densities of 
$ \sim 10^{21-22}$ cm$^{-2}$.  X-rays are a strong heating
agent in the unshielded innermost regions of the disk, but get attenuated
rapidly due to the high  midplane radial column density $N(r)$  
$\sim 10^{24}$ cm$^{-2}$ (see Eq.(\ref{col-est})). The penetration column for
2 keV X-ray photons, where the X-ray spectrum peaks, is
$\sim 1/\sigma(E) \sim 5 \times 10^{22}$ cm$^{-2}$
which implies a penetration depth of $\sim 0.003$ AU. Heating due to X-rays
becomes unimportant in the midplane for $r\gg r_i$ into the disk. X-rays, however, can 
penetrate further  at the surface of the disk, where the attenuation
columns of gas towards the line-of-sight to the star 
are low. X-rays are the dominant heating mechanism at the surface.
Gas-grain collisions are the 
strongest gas heating source through most of the equatorial regions of the 
disk even though  the dust surface area per H nucleus
is much smaller than the interstellar value(\s $10^{-3}$ times that in the ISM
for the standard case). Note that in the innermost regions,
the gas-grain collisions become a cooling mechanism, rather than a heating
mechanism, as the gas becomes hotter than the dust grains due to X-ray and
\mh formation heating.  Heating due to
cosmic rays, grain photoelectric heating and drift heating are 
negligible in this disk model. Kamp \& van Zadelhoff(2001)
find drift heating to be the dominant heating agent in their models of
gas disks around A stars, but we find that our more careful calculation
of drift velocities yields values that are typically
too low (\s $10^3$ cm s$^{-1}$ or less) for viscous heating to 
be important in our \s 20 AU disks.  Nevertheless, we 
include this mechanism as it may become important in other regions of 
parameter space. The dominant coolants in the midplane of the disk (Figure~\ref{cool})
are OH, H$_2$O, SI, SiII, \mh and FeII in the inner regions and OI, CO and OH
in the outer disk. Above the midplane SI and OI  are strong coolants.
At radii larger than $\sim 7 $ AU, the IR continuum field from the
warmer dust excites the upper level of the [SI] line, and
subsequent collisional de-excitation leads to a gas heating
rather than cooling by this line, as seen by the sharp cutoff in the
SI cooling in figure~\ref{cool}.  

The gas temperature in the midplane of the disk (Figure~\ref{tempr}) decreases
with radius approximately as $T_{gas} \propto r^{-0.8}$. 
The plot also shows the dust temperatures of the smallest 
and largest grains. These temperatures are higher than that of the gas 
over most of the disk interior and therefore gas-grain collisions heat the gas
here.  In the outer regions of the disk midplane,
the gas temperature falls well below the dust temperature, as the decreasing
density leads to a decrease in collisional coupling between grains and dust
which lowers the heating rate of the gas. Figure~\ref{tempr} also shows the
number density of gas in the midplane which falls approximately as $1/r^2$.

Figure~\ref{tempz} shows the variation of
gas temperature with height at two different disk radii, 2 and 10 AU, and the
gas number density with height at 2 AU, with the dominant
heating sources and coolants marked at various heights. In the equatorial region
where gas-grain collisions dominate the heating, the density(dashed line in
Figure~\ref{tempz}) can be seen to
decrease rapidly with $z$ (recall that for isothermal disks, $n \propto e^{-z^2/H^2}$,
where H is the scale height), leading to a reduced gas-grain coupling
and heating, and causing a drop in temperature with height.
Several scale heights above the midplane,
the attenuating column to the star drops to sufficiently low values that X-ray
heating begins to dominate and the gas temperature rapidly rises. 
The main coolant at the surface is [OI]$63\mu$m, which is optically thick 
at these column densities. The column to the surface decreases with height, and
the line becomes more optically thin and more efficient at cooling.
 This leads to a drop in temperature
again in the upper disk atmosphere. Figure~\ref{tempz} also shows the
temperature profile at a larger disk radius (10AU), where gas temperatures are
much lower. The main heating sources remain gas-grain collisions at the
midplane and X-rays at the disk surface, but the main coolants in these
regions are CO and OI, as the temperatures are too low ($< 100$ K)
 to significantly excite the 
transitions of \mha, SI, SiII or FeII, and the densities and OH and H$_2$O abundances
are too low for OH and H$_2$O to dominate the cooling. 

\begin{figure}[f]
\plotone{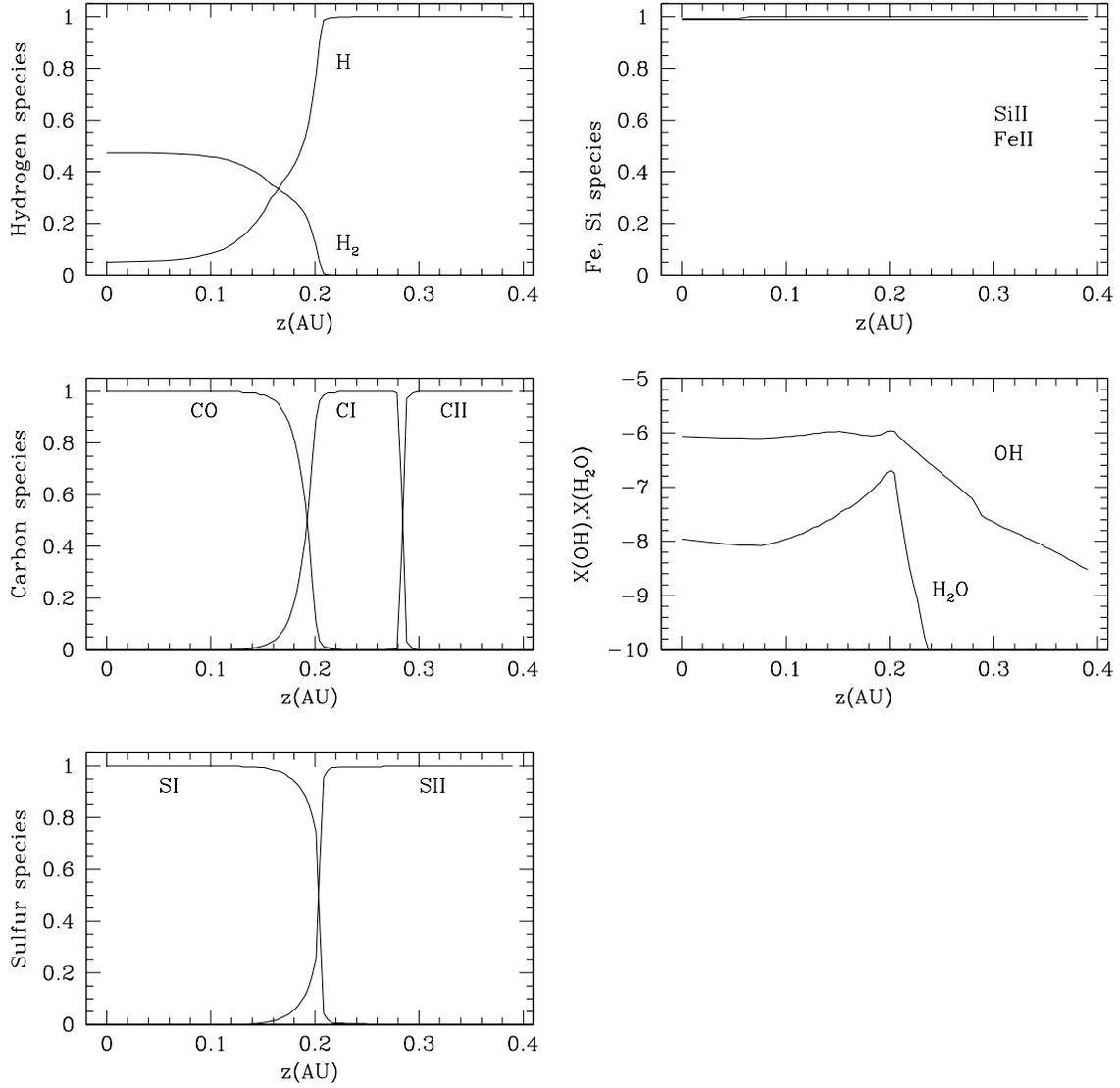}
\caption{Abundances of important species as a function of $z$ at a 
radius of 2 AU for the standard disk model. All abundances, except for
OH and H$_2$O, have been normalized to their maximum possible values.}
\label{chem}
\end{figure}

\paragraph{Chemistry}
Figure~\ref{chem} shows the
abundances of various species as a function of height above the disk
at a radius of 2 AU. We discuss below the essential species in the hydrogen,
OH/H$_2$O, C$^{+}$/C/CO, Si$^+$/Si, Fe$^+$/Fe, and S$^+$/S chemistry.

In the innermost disk, for $r\lesssim 1.1$ AU, hydrogen is predominantly
atomic. \mh is formed via reactions of H with H$^-$ and three body reactions,
but is rapidly destroyed by reactions with O to form OH and  by 
photodissociation. In regions high above the midplane, where there is 
less attenuation of stellar UV, photodissociation of \mh dominates.
Beyond $1.1$ AU, self-shielding prevents UV photodissociation of \mha.
 The destruction reaction with O has a  high activation barrier 
at $\Delta E/k = 3200$ K  and this reaction rate rapidly falls as the temperature decreases 
radially outward. Consequently, hydrogen turns fully molecular in the dense
midplane at $r \sim 2 $AU. At the surface of the disk at $r=2$ AU, 
lower attenuation columns lead
to photodissociation of \mha, and hydrogen is atomic
as is shown in Figure~\ref{chem}. 

Carbon turns molecular in the midplane even at the inner edge of the
disk where hydrogen is still atomic, and is in the form of CO
throughout the disk midplane. In the upper regions of the disk, CO
is photodissociated into atomic carbon, which gets ionized even higher
up in the disk to form C$^+$ (Figure~\ref{chem}). CO is formed via
the reaction of C with OH in the inner disk where the temperatures are high
and OH is more abundant.  O reacts with CH to form CO in the outer disk. 
The main destruction routes of CO are photodissociation by the stellar 
UV field at the inner edge of the disk and in regions above the midplane
everywhere in the disk. In the shielded interior of most of the disk,
all the available carbon forms CO, as the only destruction mechanism for
CO molecules is collisions with the very few He$^+$ ions that are formed by cosmic rays.

The gas phase elemental oxygen abundance is a factor of 2.3 higher than the carbon
abundance (Table~\ref{abun}, Appendix A) and all the oxygen not locked up in CO is
in the form of OI throughout the disk. The stellar photons are not energetic 
enough to ionize oxygen, and trace amounts of the atomic oxygen form
O$_2$, OH and H$_2$O. 
In the inner disk, where the temperature is
high (T $ > 300$ K), O reacts with \mh to form OH, which then reacts
again with \mh to form water. OH also reacts with O to form O$_2$
throughout the disk, which is destroyed by photodissociation. 
 H$_2$O and OH are
the dominant coolants at the high
densities and temperatures in the innermost disk. 
Water is photodissociated through the disk
more readily than OH as 
it has a lower photodissociation threshold (\s $ 4.3$ eV).
In disks that are more massive, larger attenuation
columns shield H$_2$O from photodissociation and due to its
high abundance it dominates the cooling, as will be discussed later.
As the temperature
drops further into the disk, the production of OH and H$_2$O drops,
because the reactions of O and OH with \mh have high activation barriers.
The resulting decrease in the abundance of OH lowers
its strength as a coolant further into the disk.  The 
destruction of OH is mainly via photodissociation which rapidly decreases
with disk radius as the optical depth in the relevant FUV band increases,
and this keeps the abundance of OH from dropping sharply. 
 The main reaction for the
production of OH and  H$_2$O in the cooler outer 
regions is recombination of  H$_3$O$^+$, which
is initiated by cosmic ray ionization of \mha.

Sulfur is predominantly atomic throughout
most of the disk, except very near the midplane where it is partly
molecular and at the disk surface where it is ionized to form SII. 
In more massive disks, which shield sulfur molecules from
photodissociation, we find that most of the sulfur is SO$_2$ 
at regions near the midplane. Silicon and iron are ionized through
most of the disk, and are neutral only in denser disks where there is
sufficient shielding from photoionization. Limited
silicon chemistry is included in our models and we find that silicon
is ionized through most of the disk. For more massive disks
with $M_{gas} > 0.1$ \mj, silicon is atomic in dense regions such as
near the midplane, but does not turn molecular even for our most massive
disks. 

SI, SiII, FeII and
\mh S(0) all have their first excitation level \s 500 K above ground and
are excited under similar conditions. The gas needs to be warm, 
$\gtrsim$ 100 K for emission from these lines to be strong. 
Note that sulfur is a much stronger coolant than \mh though both
species have similar excitation energies. At disk gas densities,
these coolants are in LTE so that their cooling is proportional to their
abundance times their spontaneous transition probability. Though the 
abundance of sulfur is 4.5 orders of magnitude lower than that of \mha, 
its  spontaneous transition probability is $4.7 \times 10^{7}$ that 
of \mh S(0). Thus, at high densities, and assuming the lines
are optically thin, [SI]25.23$\mu$m should dominate \mh 
28 $\mu$m by a factor \s 1000.  In fact, the [SI]25$\mu$m transition
is optically thick, so that the [SI]25$\mu$m line is typically
\s 10 times the \mh line strengths. Cooling from SI decreases
as the gas temperature decreases, and in the outer disk the most
important coolants are CO and OI at the midplane and  OI
 at the surface where CO is photodissociated. 

\begin{figure}[f]
\plotone{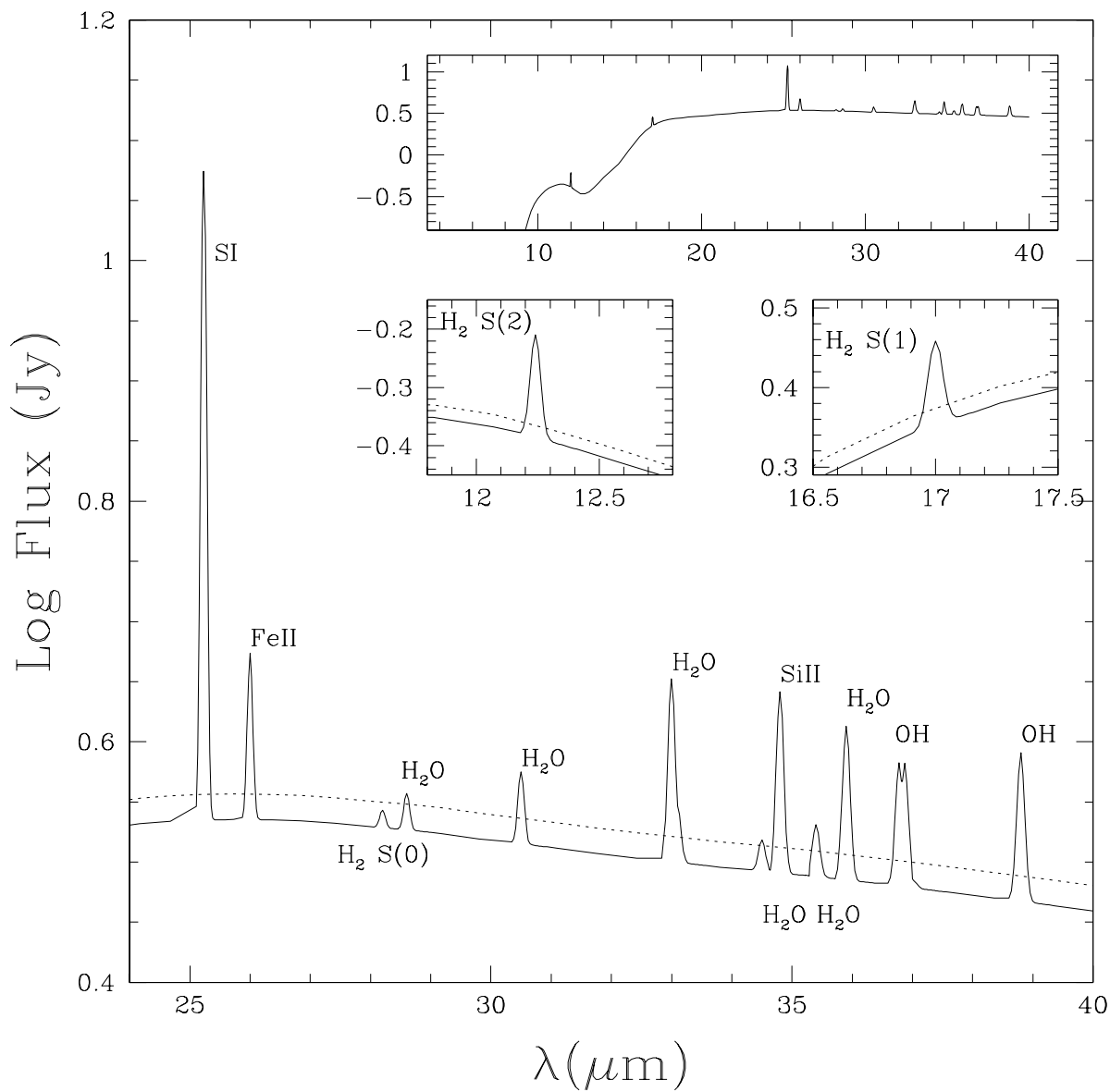}
\caption{Mid-infrared spectrum in the 24$-$40 $\mu$m wavelength region showing
the dust continuum and dominant gas emission lines for the standard disk model,
assuming a distance to the disk of 30 pc and a spectral resolving power 
$(\lambda/\Delta \lambda)$=600, typical of high resolution Spitzer
observations. The dashed line indicates a 5\% line-to-continuum ratio
above which lines may possibly be detected. The 
insets show the complete spectrum from $10-40 \mu$m, and the
S(1) and S(2) lines of \mh.}
\label{diskspec}
\end{figure}

\paragraph{Disk emission and spectrum}
Figure~\ref{diskspec} shows a model spectrum for this standard disk
at a distance of 30 pc, 
with the dominant lines in the 5$-$40 $\mu$m band, superimposed 
on the dust continuum. (The resolution of Spitzer-IRS in the high-resolution mode ($R=600$) has been
assumed  for computing the emission spectra.)
The lines shown in the emission spectrum
are strong enough for nearby sources ($\lesssim 100$ pc)
to be potentially observable by the Spitzer Space Telescope, or
future infrared telescopes, such as SOFIA or Herschel.  The IRS instrument on
the Spitzer Space Telescope  has
 a $3\sigma$ flux detection limit of \s $10^{-18}$ W m$^{-2}$, 
with an integration time of \s 500s.

The [SI]25.23$\mu$m line stands out as a strong coolant in
our disk spectrum and as an important
diagnostic of the presence of gas at \s 1 AU in disks. 
The possible detection of gas disks in sulfur
is one of our main results, as in many cases this line dominates the spectrum
and is stronger than the H$_2$ lines. The [OI]63$\mu$m
and [SI]56.0$\mu$m lines may also be strong enough for detection by SOFIA. 
Other detectable lines in this
figure are the \mh S(1) and S(2) lines, the 
[FeII]26.0$\mu$m line, the
[SiII]34.8$\mu$m line, and water lines at 
33$\mu$m (o-H$2$O, $6_{6,1} \rightarrow 5_{5,0}$) 
and 36$\mu$m (o-H$2$O, $6_{5,2} \rightarrow 5_{4,1}$).

\begin{figure}[f]
\plotone{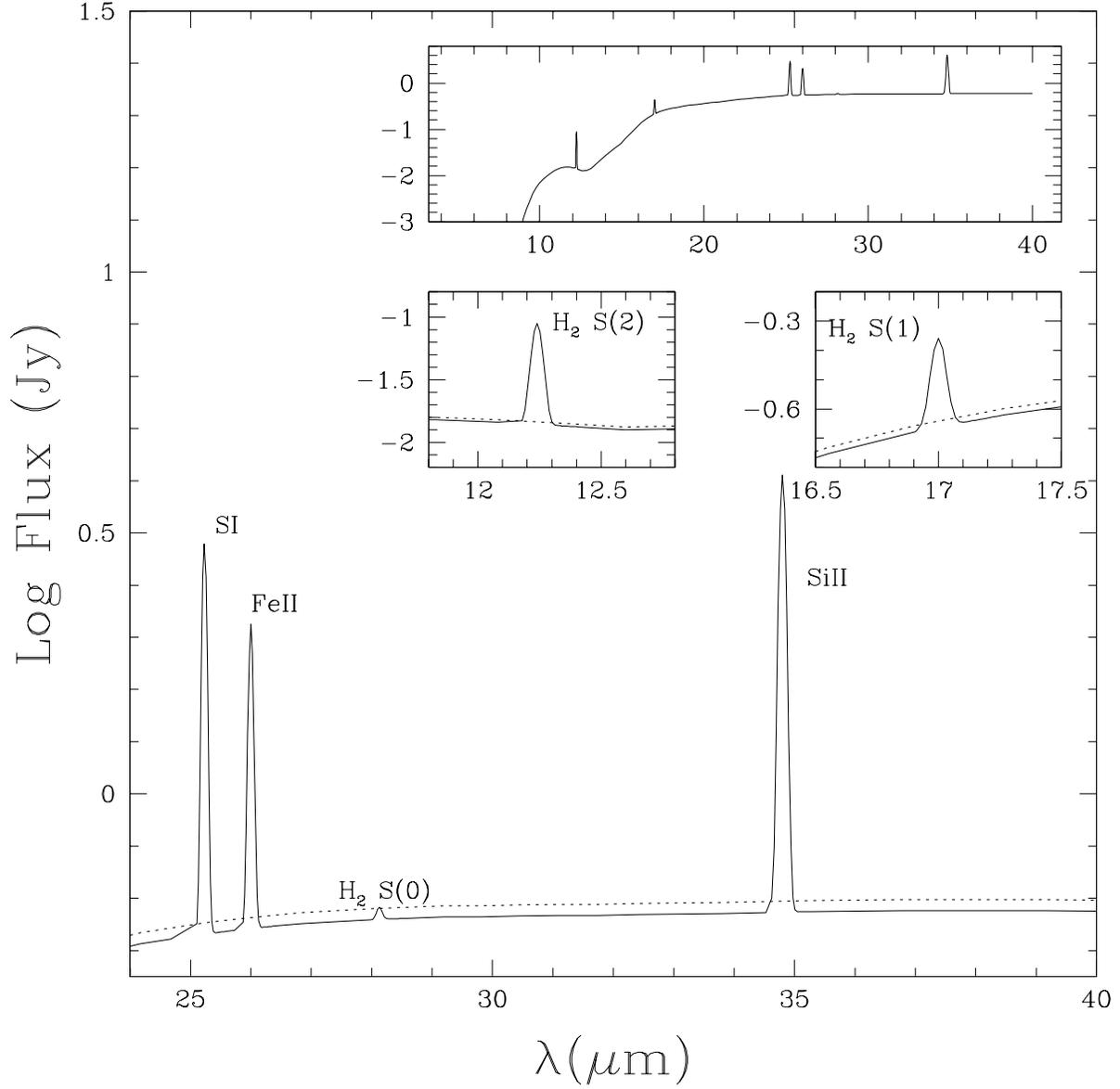}
\caption{Mid-infrared spectrum (as in Figure~\ref{diskspec}), for
a disk with an inner radius of 10 AU,  and with all other parameters 
the same as the standard model.}
\label{diskspec10}
\end{figure}

Figure~\ref{diskspec10} shows a similar spectrum for a disk with
an inner radius of 10 AU, which is seen to have no strong OH or
H$_2$O lines, but has detectable \mh S(1), \mh S(2),
[SI]$25.23\mu$m, [FeII]$26\mu$m and [SiII]$34.8\mu$m lines. The
effect on the gas of changing the inner disk radius is discussed in 
more detail in \S 5.2. 

Table~\ref{linelist} lists the observable mid-infrared 
and far infrared lines  with their wavelengths and the 
telescope instruments that can observe them.   
\begin{table}[f]
\caption{Infrared Emission Lines, their Wavelengths and
Observing Instruments.}
\label{linelist}
\begin{tabular} {l r c l}
\hline
Species & $\lambda (\mu$m) &  Transition & Telescope instruments \\
\hline
\mh S(2) & 12.3 & & Spitzer IRS, SOFIA EXES \\
\mh S(1) & 17.0 & & Spitzer IRS, SOFIA EXES \\
\mh S(0) & 28.2 & & Spitzer IRS, SOFIA EXES \\
OH & 15-200 & & Spitzer IRS,SOFIA EXES, GREAT, SAFIRE, Herschel SPIRE \\
H$_2$O  & 15-200 & & Spitzer IRS,SOFIA EXES, GREAT, SAFIRE, Herschel SPIRE  \\
FeI & 24.0 & $^5$D$_3-^5$D$_4$  & Spitzer IRS, SOFIA EXES \\
SI & 25.2 & $^3$P$_1-^3$P$_2$ & Spitzer IRS, SOFIA EXES \\
FeII & 26.0 & $^6$D$_{7/2}-^6$D$_{9/2}$ & Spitzer IRS, SOFIA EXES \\
SiII & 34.8 &  $^2$P$_{3/2}-^2$P$_{1/2}$& Spitzer IRS  \\
FeI & 34.2 &  $^5$D$_2-^5$D$_3$& Spitzer IRS \\
FeII & 35.4 & $^6$D$_{5/2}-^6$D$_{7/2}$ & Spitzer IRS \\
SI & 56.6 & $^3$P$_0-^3$P$_1$ & SOFIA FIFI-LS \\
OI & 63.1 &  $^3$P$_1-^3$P$_2$& SOFIA FIFI-LS, GREAT, Herschel PACS \\
OI & 145.6 &  $^3$P$_0-^3$P$_1$& SOFIA SAFIRE, GREAT, Herschel SPIRE \\
CII & 157.7 & $^2$P$_{3/2}-^2$P$_{1/2}$ & SOFIA SAFIRE, GREAT, Herschel SPIRE\\
CI & 369.0 & $^3$P$_2-^3$P$_1$ & SOFIA CASIMIR, SAFIRE, Herschel HIFI\\
CI & 609.2 & $^3$P$_1-^3$P$_0$ & SOFIA SAFIRE, Herschel HIFI\\
\hline
\end{tabular}
\end{table}
Gas emission lines are potentially important diagnostic 
tools of physical and chemical conditions of gas in disks and could 
be useful in detecting even very small
amounts of gas. However, most of these lines are optically thick with the
exception of H$_2$ S(0) and would correspondingly diminish in strength with the
inclination of the disk, being the strongest for a face-on orientation.
Molecular hydrogen remains optically thin 
in the S(0) $28\mu$m line  and the line strength is not
affected by the orientation angle of the disk. Moreover, it is less dependent
on the chemical network and the cosmic abundances assumed and directly 
traces the gas in the disk.  However,  the
\mh S(0) line is weaker than other gas emission lines and  hence is
 more difficult to detect. For pure sensitivity to the presence of small amounts of gas at 
$r \sim 1-2 $ AU, the [SI]25.23$\mu$m line is predicted to be the best probe. 

\section{Parameter Survey and Emission Line Detectability}
\subsection{Variation of Spectra with M$_{gas}$ and M$_{dust}$.}
\begin{figure}[f]
\plotone{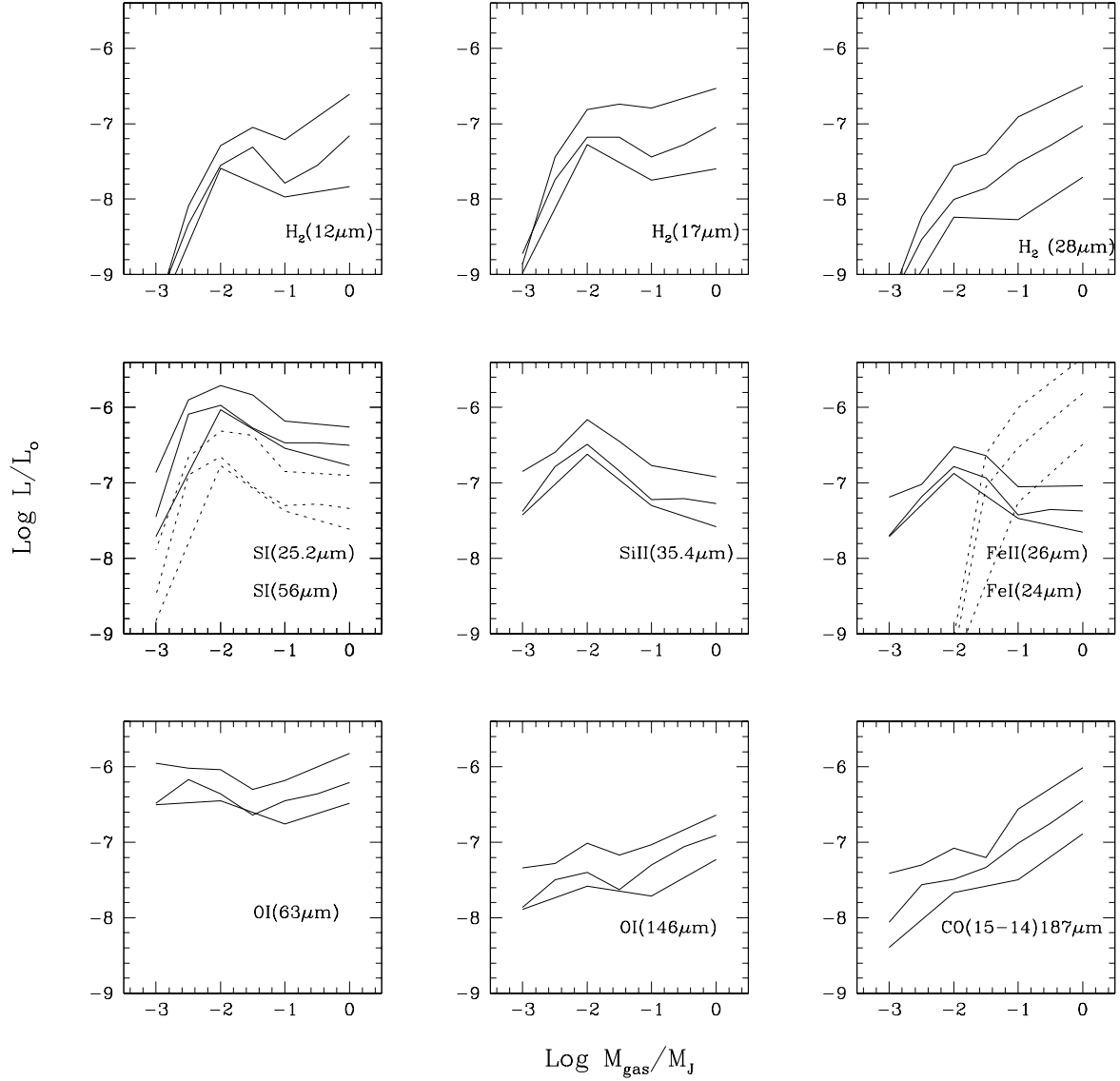}
\caption{Mid-IR line luminosities as a function of disk gas mass, for
different dust masses ($10^{-5}$ \mj\ upper line, $10^{-6}$ \mj\
middle line, and $10^{-7}$ \mj\ lower line) in the disk. The dashed
lines show the luminosities of the [SI]56$\mu$m line
and the [FeI]24$\mu$m line.}
\label{vmass}
\end{figure}

\begin{figure}[f]
\plotone{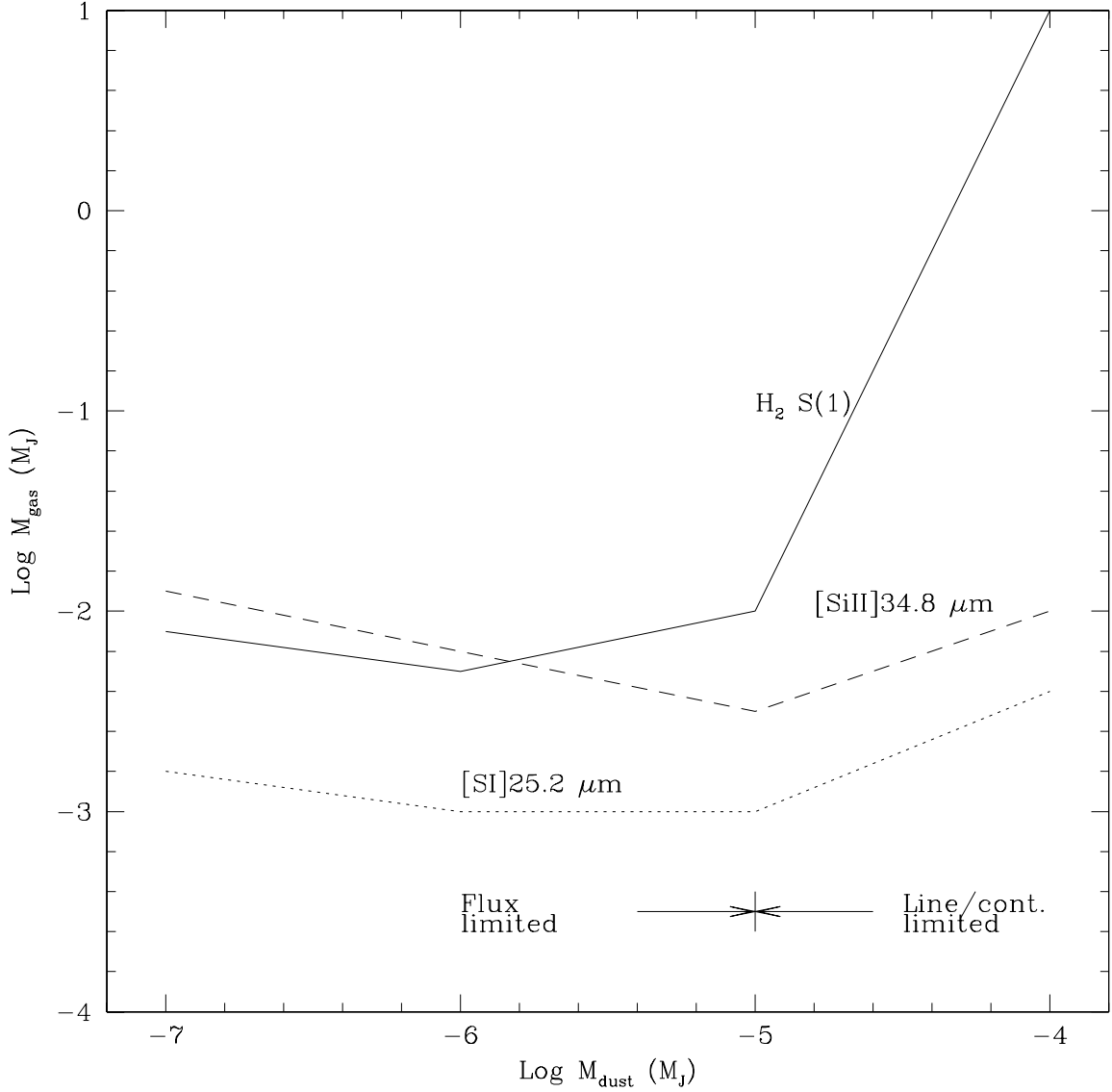}
\caption{Detectability plots for three representative gas
emission lines \mh S(1), [SI]25.23$\mu$m and [SiII]
34.8$\mu$m, by the Spitzer Space Telescope (IRS instrument with R=600)  
for a standard model disk at 30 pc. The lines 
show the minimum detectable gas mass for different dust
masses in the disk for each line. For M$_{dust} < 10^{-5}$
\mj, detection is flux limited ($3 \times 10^{-8}$ L$_{\odot}$ 
for $\lambda \sim 25 \mu$m).
For M$_{dust} > 10^{-5}$ \mj, the line-to-continuum ratio
is the limiting factor. We assume that a line-to-continuum ratio
of 5\% is sufficient for detection. }
\label{detect}
\end{figure}

We have conducted a parameter survey to
explore  the probability of gas emission line detection above the dust continuum
emission for a range of gas masses and dust masses and for a range of 
 stellar and disk parameters (Table~\ref{range-par}).
\begin{table}[f]
\caption{Parameter Range for Disk Model}
\label{range-par}
\small
\begin{tabular}{l r l r}
\hline 
Gas mass & $10^{-3}-1$ M$_J$ & Dust mass & $10^{-7}-10^{-4}$ M$_J$ \\
Inner radius $r_i$ & 0.3-10 AU  & Minimum grain size $a_{min}$ & 0.1-10.0 $\mu$m \\
Outer radius $r_o$ & 20 AU   & Maximum grain size $a_{max}$ & 1.0 mm\\
Surface density power law index $\alpha$ $$ & 0.5-2.0 & Grain size dist. index $s$
& 3.5 \\
Stellar X ray Luminosity & $10^{-2}-10^{-5}$ L$_{bol}$ & 
Stellar temperature $T_*$ & 4500K and 5780 K \\
\hline
\end{tabular} 
\end{table}
We limit the range of parameters to those
which might give us detectable infrared line fluxes, and those whose
dust optical depths $\tau_V \lesssim 5 $. 
Figure~\ref{vmass} shows the change in IR line luminosities due to variations
in gas and dust masses. 
For a disk at a distance of 30 pc, the flux detection limit
of the IRS (high resolution) instrument on Spitzer 
for a $3\sigma$ 500-second detection corresponds to a line
luminosity of \s $3 \times 10^{-8}$ L$_{\odot}$ (or $10^{-18}$
W m$^{-2}$). 
For comparison, SOFIA at these wavelengths (10-28 $\mu$m) 
has a minimum detectable
$3\sigma$ line flux of  \s $10^{-17}$ W m$^{-2}$ 
($3 \times 10^{-7}$ L$_{\odot}$), and
Herschel \s $7 \times 10^{-18}$ W m$^{-2}$ ($2 \times 10^{-7}$ L$_{\odot}$)
at its shortest observing
wavelengths (60-200 $\mu$m). Instruments on both telescopes have resolving
powers \s $10^{3-4}$. 
Figure~\ref{detect} presents these results in an alternate way,
which readily shows the parameter space where lines are detectable.

The S(1) and S(2) lines of \mh arise from gas with temperatures
$\gtrsim$ 150-200 K, whereas S(0) emission has a significant
contribution from slightly cooler gas ($T \gtrsim 100 K$). At very
low gas masses ($10^{-3} $ \mj), the midplane column density through the
disk  is $\sim 10^{23}$ cm$^{-2}$, and hence low enough to
allow penetration by stellar UV and X-ray photons through much of the
inner disk and to photodissociate \mha. The low amount of warm \mh
gas results in very low line strengths of all three \mh lines, as
can be seen in Figure~\ref{vmass}. As the disk gas mass is increased
to $10^{-2}$ \mj, the column density increases allowing shielding
of \mha, increasing its abundance in the inner, warm regions of
the disk. The gas in the inner disk where the emission from the S(1)
and S(2) lines peak, is heated by collisions with dust,
\mh formation heating and by X-rays. The increased amounts of warm
\mh gas ($\gtrsim 100 $ K) result in a sharp increase in the line strengths. 
As the gas mass is further increased to 0.1 M$_J$, the gas temperature at a
fixed radial column density initially decreases because the H$_2$O cooling rises
faster than the X-ray heating with density. Thus, there is less mass and
surface area of warm gas. As the gas mass increases to $\sim 1$ M$_J$, the
gas heating by gas-dust collisions takes over and raises the gas temperature
and the S(1) and S(2) luminosities.  The S(0)
line emission peaks at regions slightly further out where dust
dominates the heating and it monotonically increases with gas mass.
At low gas masses, the observed \mh S(0) emission directly correlates 
with the mass of {\em warm} gas in the
disk, regardless of the dust or total gas masses, and  the warm gas 
(T $\gtrsim$ 100 K) mass in a disk at 30 pc 
needs to be above \s $10^{-3}$ \mj\ for detection by the Spitzer Space Telescope. 
 At higher gas masses, \s 1 \mj, dust collisions are the main
heating mechanism that heat \mh gas.  In these relatively
massive disks, \mh line emission does not
increase proportionately with mass, as the lines start to become
optically thick (at disk gas masses $\gtrsim 0.05$ \mj\ for S(1) and
S(2), $\gtrsim$ 0.2 \mj\ for S(0)). 
Figure~\ref{vmass}
shows that for line luminosities $\gtrsim 5 \times 10^{-8}$ L$_{\odot}$, the luminosity
generally scales with a low power($<1$) of gas mass. This means that for disks at
distances $d\gg30$ pc, considerably more gas mass is required for detection
than if $L\propto M_{gas}$ is assumed. This important point will
be discussed more in \S 6. 

An increase in total dust mass of the disk produces increased emission
due to an increase in  heating of the gas via gas-dust collisions.
Dust collisions are an important heating mechanism 
in the bulk of the \mh emitting regions. 
As the dust mass is decreased, the decreased coupling between gas and dust lowers
the gas heating rate and the gas temperature and therefore the emission. 
  On the other hand, dust produces IR continuum emission
which can obscure the detection of infrared lines (typically,
systematics introduce baseline variations which make it impossible to detect
lines at the $\lesssim$ few per cent continuum level). There
is thus an optimal gas-to-dust mass ratio $\sim 10^{3} - 
10^{5} $ and gas mass $\gtrsim 10^{-2} $\mj\ (for disks at 30 pc) 
where H$_2$ emission can be detected. 
Given $\gtrsim  10^{-2}$ \mj\ of total gas mass, a gas to dust mass ratio
$\lesssim$ 100 produces a line to continuum ratio of $\lesssim 5 \%$ for the
highest Spitzer Space Telescope resolving power (R=600),
rendering the line difficult to detect above the strong continuum.
 On the other hand, if the
dust abundance is lowered to the point where the gas to dust ratio $\gtrsim 10^5$,
then the grain-gas collisional heating is so drastically reduced that the
gas is cold, and the \mh lines are too weak to detect in the absence of
other gas heating mechanisms.

The SI, FeII, and SiII emission line strengths are  complicated by 
chemistry in the disk.
For disks of low gas mass ($10^{-3}-10^{-2}$ \mj),
the main sulfur species in the denser regions near
the midplane of the disk is atomic sulfur, and sulfur is photoionized near the surface. Iron and
silicon are completely ionized throughout such disks.
  In this low gas mass limit, the emission increases as the gas mass
increases and as the dust mass increases.
A higher dust mass, in general, implies more heating through gas-grain collisions
and hence warmer gas. As the gas mass increases to 0.1 \mj \ or higher,
the main sulfur-bearing species near the midplane 
is SO$_2$, as sulfur  turns molecular at
these high densities and opacities. Here most of the atomic sulfur is in a layer above the midplane
where there is enough penetration of stellar radiation to photodissociate SO$_2$, but
not sufficient to ionize sulfur.  Similarly, chemistry also affects the
FeII and SiII lines, as there is substantial amounts of neutral iron and silicon 
in the denser regions (i.e. near the inner midplane)
of the more massive disks. The SI line is optically thick for all
the gas masses considered here and at higher gas masses, the line
luminosity is approximately independent of mass. There is a slight 
increase in the  line strength at $10^{-2}$ \mj\ as X-rays also contribute
to the gas heating making the gas warmer,
and a sharp decrease at lower gas masses as sulfur
gets ionized due to lower column densities. The [SI]25.23$\mu$m line is stronger than the
56 $\mu$m line by a factor of $3-10$ in all cases. The SiII and
FeII line luminosities show  a similar dependence with gas mass, becoming independent of mass
the lines get optically thick for gas masses $\gtrsim 10^{-2}-0.1$ \mj. 
The emission from [FeI]24$\mu$m, on the other hand, increases rapidly
with gas mass ($\gtrsim 0.1$ \mj) due to the increased abundance of warm 
dense FeI gas. The [FeI]24$\mu$m line is the strongest line, with luminosities
exceeding that of [SI]25.23$\mu$m for massive disks, making it an excellent
tracer of gas in disks with masses $\gtrsim$ 0.1 \mj. However, its line
strength depends on the gas phase abundance of Fe in these disk regions,
which is uncertain. 

The [OI]63$\mu$m line is strong, and does not show much variation
in strength with increasing gas mass, reflecting the fact that the 
[OI] line is nearly always optically thick. The
emission is only weakly dependent on the dust mass, as it arises 
 mostly from the surface regions of the disk where X-rays dominate the
heating. The [OI]146$\mu$m line is weaker than the 63 $\mu$m line by
approximately a factor of 10, but is strong enough to be detected in
more massive (and more dusty) disks. 
Emission from the [CII]158$\mu$m line from $r\le20$ AU is too weak to be observed by Herschel
or SOFIA, as the region of ionized carbon at the disk  surface is too
low in mass to be significant.  CO emission increases with increasing gas mass.
CO emission is optically thick for low J transitions
and is dependent on the temperature and area of the emitting surface.
The line strength increases for moderately high J transitions, whose
wavelengths lie closer to the blackbody peak of the warm disk gas,
and peaks at the optically thin  $15\rightarrow14$ transition at 187$\mu$m
which arises mostly from within \s 10 AU of the disk.

Emission from various lines of H$_2$O and OH is found to be strong for disks
with gas masses $\gtrsim$ 0.1 \mj. In these disks, the densities
are high and the abundance of these species is high making them strong
coolants. OH is a strong coolant when the gas temperature is high (\s few
100 K), and OH line emission mostly arises from the dense, hot inner disk
where H$_2$O photodissociates to OH more readily.  Cooling by H$_2$O  dominates in the
dense equatorial regions of these massive disks at all radii, 
but the emission is spread in various lines in the mid-IR,
only some of which lie in the the Spitzer Space Telescope  band from 10$-$37 $\mu$m. The abundance
of lines also implies that the cooling luminosity is distributed over many
lines which reduces the strength of each individual line. Table~\ref{water}
lists the strongest infrared line luminosities of the ortho and para-H$_2$O species and OH 
for a disk with a gas mass of $0.1$ \mj\ and dust mass $10^{-5}$ \mj,
with all other parameters having their fiducial values. 
\begin{table}[f]
\caption{Line luminosities of OH and H$_2$O lines for a disk with a gas
mass of 0.1 \mj\ and dust mass $10^{-3}$ \mj, with all other
parameters at their fiducial values.}
\label{water}
\begin{tabular}{lrrlrr}
\hline
Species & $\lambda (\mu$m) & Line lum. (in L$_{\odot}$) & 
Species & $\lambda (\mu$m) & Line lum. (in L$_{\odot}$) \\
\hline
OH &  30.54 & 9.8$\times 10^{-8}$ & OH &  30.63 & 1.0$\times 10^{-7}$\\
OH &  33.80 & 1.4$\times 10^{-7}$ & OH &  33.93 & 1.4$\times 10^{-7}$\\
OH &  37.01 & 1.8$\times 10^{-7}$ & OH &  37.08 & 1.8$\times 10^{-7}$\\
OH &  38.84 & 7.7$\times 10^{-7}$ & OH &  38.88 & 6.5$\times 10^{-7}$\\
o-H$_2$O  &  21.15 &   2.6$\times 10^{-7}$ & 
p-H$_2$O  &  21.17 &  2.6$\times 10^{-7}$\\
p-H$_2$O  &  21.18 &  3.0$\times 10^{-7}$ & 
p-H$_2$O  &  21.83 &  2.8$\times 10^{-7}$\\
o-H$_2$O  &  21.85 &   3.1$\times 10^{-7}$ & 
p-H$_2$O  &  21.89 &  2.7$\times 10^{-7}$\\
p-H$_2$O  &  23.19 &  3.0$\times 10^{-7}$ & 
o-H$_2$O  &  23.90 &   3.2$\times 10^{-7}$\\
o-H$_2$O  &  25.17 &   2.9$\times 10^{-7}$ & 
p-H$_2$O  &  25.22 &  3.1$\times 10^{-7}$\\
o-H$_2$O  &  25.94 &   4.0$\times 10^{-7}$ & 
p-H$_2$O  &  25.98 &  3.1$\times 10^{-7}$\\
p-H$_2$O  &  27.03 &  3.1$\times 10^{-7}$ & 
o-H$_2$O  &  27.99 &   3.1$\times 10^{-7}$\\
o-H$_2$O  &  28.59 &   2.7$\times 10^{-7}$ & 
p-H$_2$O  &  28.59 &  2.7$\times 10^{-7}$\\
p-H$_2$O  &  28.91 &  2.3$\times 10^{-7}$ &
p-H$_2$O  &  29.36 &  2.7$\times 10^{-7}$ \\
o-H$_2$O  &  29.84 &   3.0$\times 10^{-7}$ & 
p-H$_2$O  &  29.88 &  2.9$\times 10^{-7}$\\
p-H$_2$O  &  30.47 &  2.6$\times 10^{-7}$ & 
p-H$_2$O  &  30.53 &  2.7$\times 10^{-7}$\\
o-H$_2$O  &  30.53 &   2.8$\times 10^{-7}$ & 
o-H$_2$O  &  30.87 &   2.5$\times 10^{-7}$\\
o-H$_2$O  &  30.90 &   3.4$\times 10^{-7}$ & 
p-H$_2$O  &  31.74 &  2.5$\times 10^{-7}$\\
p-H$_2$O  &  33.00 &  2.9$\times 10^{-7}$ & 
o-H$_2$O  &  33.01 &   3.0$\times 10^{-7}$\\
p-H$_2$O  &  34.40 &  2.0$\times 10^{-7}$ & 
o-H$_2$O  &  34.55 &   2.7$\times 10^{-7}$\\
o-H$_2$O  &  35.43 &   2.7$\times 10^{-7}$ & 
p-H$_2$O  &  35.47 &  3.2$\times 10^{-7}$\\
o-H$_2$O  &  35.67 &   2.2$\times 10^{-7}$ & 
p-H$_2$O  &  35.90 &  2.9$\times 10^{-7}$\\
p-H$_2$O  &  36.21 &  2.7$\times 10^{-7}$&
o-H$_2$O  &  66.44 &    2.6$\times 10^{-7}$\\
o-H$_2$O  &  75.38 &    3.6$\times 10^{-7}$&
p-H$_2$O  & 100.98 &   8.1$\times 10^{-7}$\\
o-H$_2$O  & 108.07 &    3.3$\times 10^{-7}$&
p-H$_2$O  & 138.53 &   5.9$\times 10^{-7}$\\
o-H$_2$O  & 179.53 &    2.2$\times 10^{-7}$ \\
\hline
\end{tabular}
\end{table}

\subsection{Variation of Spectra with Other Parameters.}
The position of the inner radial cutoff, $r_i$ is an important parameter
which can alter the observed strength of emission lines. We have run cases with
four different values of this inner radius equal to 0.3, 1.0, 3.0 and 10 AU from the
central star, keeping all other parameters identical to our standard disk model.
Disks with smaller inner holes have 
a small mass of hotter, denser gas which lies closer to the
central star, and might be  expected to have increased line luminosities as compared to
disks with large inner holes. However, larger inner holes increase the gas mass
heated and chemically altered by X-rays and UV, so the
situation is in fact more complicated. 
The fluxes for all the lines (except FeI)  are seen to drop when the inner radius is
decreased from 1 AU to 0.3 AU. The higher disk densities
in the inner disk cause sulfur to turn molecular, 
and reduce the ionization of iron and silicon. With lower abundances
of S, Fe$^{+}$ and Si$^{+}$ and a smaller surface area of warm
gas, their line luminosities correspondingly decrease. The
line luminosity from the FeI line however increases due 
 to a higher abundance of neutral iron. The
strength of the \mh lines is also seen to decrease slightly, and this is due to 
lower masses of $\gtrsim$ 100 K \mh gas for the $r_i=0.3 $ AU disk because of
 effective self-shielding from UV and X-rays. When the disk inner radius is increased
from 3 AU to 10 AU, emission from all the lines except SI and \mh S(0)
increases again. A disk with such a large inner radius has a significantly low
radial column density ($\sim 4 \times 10^{23}$ cm$^{-2}$ through the
midplane) so that X-rays  more effectively penetrate the disk,
and now dominate  gas-grain collisions as the main heating mechanism. The
enhanced emission in most of the lines is due to a higher gas temperature due
to X-ray heating through most of the disk. Emission from the [SI] lines
is seen to decrease, however, and this is because of a higher ionization
rate of sulfur which lowers the atomic sulfur abundance and lowers the effective
area of warm [SI] gas relative to a disk with a smaller inner disk radius.

Variations in the surface density power-law profile (Figure~\ref{valpha}) have similar
effects on the line intensities as changing $r_i$. The amount of warm
gas in the inner regions is sensitive to the power-law index $\alpha$, and
for a steeper power-law there is more mass in the inner region as compared
to a shallower surface-density profile. We have considered three different
power-law indices, $\alpha=0.5, 1,$ and $ 1.5$. For the largest value of $\alpha$,
there is a decrease in the strength of the [SiII]34.8$\mu$m and [FeII]26$\mu$m
lines, as the increased densities in the inner regions increase the shielding
of these species from photoionization, and their abundances  decrease.
Correspondingly, the [FeI]24$\mu$m line increases sharply for large $\alpha$. 

The X-ray luminosity of the star was taken to be $10^{-4}$ times the
bolometric luminosity in the standard case, as is typical for a weak-lined
T Tauri star or more evolved stars (Feigelson et al. 2003).
However, this value may be higher
or lower depending on the age of the star and other factors, and we have run cases where we
assume an X-ray luminosity of $10^{-5}$ to $10^{-3}$ times the bolometric luminosity.
The X-ray spectrum is assumed to remain the same, peaking at around 2 keV
and ranging from 0.5$-$10 keV. The X-ray luminosity has surprisingly little
effect on the Spitzer Space Telescope  lines as can be seen from Figure~\ref{vLx}. The line luminosities
seem largely insensitive to the amount of X-ray flux, and this can be understood
by referring to the temperature structure (Figures~\ref{tempr} and \ref{tempz}) 
for the standard case. 
The X-rays only penetrate and heat a very small fraction
of the mass of the disk, and in these regions, most of the heating is carried
away as infrared emission from dust and by the [OI]63$\mu$m line. 
The strength of this line correlates
with the X-ray flux as it comes mostly from the upper surface regions of the
inner disk where the gas is warm and heated by X-rays (see Figure~\ref{tempz}).

The infrared line emission from gas is quite sensitive to the stellar flux,
with stronger emission from gas disks around more luminous stars. We have run
our disk models for two different stellar types, a G2 star and a K4 star.
In both cases, all disk parameters were the same, and the X-ray 
luminosity in both cases was $10^{-4}$ the stellar luminosity. With lower
X-rays and  UV flux and with the dust
grains around the less luminous K star being colder,  the gas 
temperature is much lower in the disk around the K4 star. 
Consequently, the line emission strengths
are correspondingly lower (Figure~\ref{vstar}). The [SI]25.23$\mu$m,
[SiII]34.8$\mu$m, [FeII]26$\mu$m
and the [OI]63$\mu$m lines may still be  strong enough for detection
by the Spitzer Space Telescope and SOFIA for disks at $d\lesssim30$ pc,
especially if the X-ray luminosities are higher than average.

Lowering the minimum grain size in the dust distribution increases line
emission, as can be seen from Figure~\ref{vamin}. In our standard case, the
minimum grain size, $a_{min}$ was chosen to be 1 $\mu$m, and we have run cases where
the minimum grain size was 0.1 $\mu$m and 10$\mu$m. Smaller dust grains are less
efficient at re-radiating stellar radiation, and get hotter as the grain size
decreases. In addition, for a fixed dust mass and $s=3.5$, the grain surface area
is proportional to $a_{min}^{-1/2}$, and therefore decreasing $a_{min}$ increases
the surface area. Both of these effects raise the gas temperature, since the
gas heating is largely by gas-grain collisions and this heating rate
is proportional to $(T_{dust}-T_{gas})$ and the dust surface area. The resultant higher gas
temperatures raise the total infrared line emission from the disk.

 In all our models,
we did not allow the formation of \mh on dust grains. It is not clear if hydrogen 
atoms stick on warm dust grains long enough to allow the formation of \mh (Cazaux \&
Tielens 2002) and we have run the standard model where we allow the grain formation of
\mha. We find that there are no significant differences in the 
resultant spectrum of the disk in the two cases. Even when
grain formation of \mh is set to zero, \mh is formed  efficiently 
by reactions with H$^{-}$ and by 3-body reactions. 

\section{Discussion, Summary and Conclusions}
We have modeled the \s 0.3 AU to 20 AU regions of ``intermediate-aged''
and ``debris'' disks
with a range of dust masses $10^{-7}-10^{-4}$ \mj (or $f_D=
L_{IR}/L_{bol} = 10^{-4} - 10^{-1}$) and gas masses $10^{-3}-1$
\mj. We define dust as grains with size $a\le 1$ mm,  and, 
assuming a size distribution $n_{dust}(a) \propto a^{-3.5}$, we
find that these disks are reasonably optically thin to starlight for 
$M_{dust} < 10^{-4}$ \mj. Therefore, we have covered all of parameter space
relevant to intermediate-aged and debris disks with sufficient gas and dust to be detected in IR
emission lines in the next decade by observatories such as the Spitzer
Space Telescope, SOFIA and Herschel. 

Although the dust is optically thin to stellar photons, the
gas is optically thick at UV and X-ray wavelengths. Therefore, we find that
the dominant gas heating mechanism for the bulk of the gas 
which is shielded from heating by UV or X-rays by gas in the inner regions
or at the surface, is gas-dust
collisions. In these regions the dust is heated by stellar
photons and colder gas is heated by collisions with warmer dust grains.
However, at the surface of the disk and at the inner edge of the disk (assuming
a hole or a gap between disk and star), the UV and X-rays heat the gas, and the 
gas temperature can exceed the temperature of even the smallest grains. 
We find that the gas temperature in the 0.3 AU to 20 AU regions of debris
disks range from \s 1000K to \s 50 K. 

The chemistry in the innermost regions and at the surface of the disk
is dominated by photoionization and photodissociation caused by UV and X-rays.
However, in the cooler, shielded regions the chemistry is dominated, 
like interstellar cloud chemistry, by ion-neutral reactions with ionization
by cosmic rays, UV and X-rays. We  assume that at $r\lesssim 20 $ AU,
the grains are too warm ($T_{dust}\gtrsim 100$ K) to support icy mantles or to 
allow surface catalysis of \mha. However, \mh is still efficiently
formed by the reaction of H with H$^{-}$ and by three-body reactions. In  warmer 
regions ($T_{gas}\gtrsim 300$ K) neutral-neutral reactions with activation
barriers become important, with the general result of enhanced production of OH
and H$_2$O. In addition, the reaction of O with \mh to form OH can dominate the 
destruction of the \mh molecule. Photodissociation of molecules by UV photons
play a large role in molecular destruction and creation of ions such as
Fe$^{+}$ and Si$^{+}$. We find that sulfur is S$^+$ at the disk surface, S at
intermediate gas opacities, and SO$_2$ in regions well shielded by overlying
gas opacity. Because \mh and CO so readily self-shield and are shielded by
atomic C, we find that for cases with sufficient gas mass to be detected
($M_{gas} \gtrsim 10^{-2}$ \mj) the bulk of hydrogen is \mh and the bulk
of carbon is CO in the disks. Unlike the case of interstellar clouds, the
carbon becomes CO before the hydrogen becomes \mh as one moves to greater depths
from the disk surface, a result caused by the lack of production of \mh on
grain surfaces in the warm disk regions. 

Surprisingly, we find that the strongest mid-IR line is [SI]25.23$\mu$m. This
results from the high value of the product of its abundance times its spontaneous decay rate
(or A-value). 
We find that [SI] is a strong
coolant in the inner disk for all our disk models with various input parameters. We have included
detailed sulfur chemistry to determine the abundance of atomic sulfur in the
disk, and emission is strong enough for detection even when substantial amounts
of gas has turned molecular (mainly SO$_2$). The total mass of gas in disks is a little more
difficult to obtain from measurements of the [SI]25.23$\mu$m line alone, as
the amount of neutral atomic sulfur is affected by the chemistry of the disk. 
In addition, the [SI] line tends to be optically thick, and therefore
not sensitive to the total gas mass. 

Other strong mid-IR lines include [SiII]34.8$\mu$m, [FeII]26$\mu$m, [FeI]24$\mu$m
and OH and H$_2$O rotational transitions.
We assume depleted gas-phase silicon and iron abundances (see Table~\ref{abun},
Appendix A)
similar to ice-free regions of the ISM.  However, we have no iron chemistry
in our models other than ionization and recombination.
We do include more chemistry for silicon and 
consider some molecular species of Si, such as SiO and SiH$_2$.
However, the abundances of the silicon molecules are calculated to be very low,
primarily because of photodissociation. 
In disks with low gas masses ($M_{gas} \lesssim 10^{-2}$ \mj), Si and
Fe are ionized through most of the disk, and the line intensities we compute are
probably reasonably accurate even though we do not treat the chemistry of these
species in detail.  In disks with higher gas masses, Si and Fe are
shielded from photoionization and the [FeI]24$\mu$m line becomes strong. 
The OH and H$_2$O lines become detectable at relatively
high gas masses, $M_{gas} \gtrsim 10^{-2}$ \mj. 

Pure rotational \mh lines such as S(0) 28 $\mu$m
and S(1) 17 $\mu$m are roughly an order of magnitude weaker than the [SI]25.23$\mu$m
line. The highest spectral resolving power of the Spitzer Space Telescope is R\s 600,
and this modest resolution makes it difficult to detect weak mid-IR
(10$-$37 $\mu$m) lines in the presence of
strong(dust) mid-IR continuum. Detection of the relatively weak \mh lines above the dust
continuum requires a relatively large ratio ($\gtrsim 1000$) of gas mass to dust mass.
Such large ratios are possible at a transition epoch (\s $10^7$ years) where
much of the dust mass has coagulated into particles of sizes $a\gg 1$ mm, but the 
gas has not yet appreciably dispersed.

Most of the Spitzer Space Telescope high spectral resolution emission
arises from gas at $r\lesssim 5 $ AU in our models, whereas
longer wavelength (lower excitation) emission such as from low to mid-J CO lines
and OI 63 $\mu$m has significant contribution from gas at $r\gtrsim 5 $ AU.
The OI line arises from the entire disk and is 
 a good probe of the conditions throughout the disk.
While the mid-IR lines observable by the Spitzer Space Telescope  are insensitive 
to the strength of the X-ray flux, the intensity
of the [OI]63$\mu$m line does increase with the X-ray luminosity. SOFIA is
capable of observing this transition, and may be able to detect smaller amounts
of gas in disks around stars that are strong X-ray sources. Our models 
indicate that CO emission is strong enough to be detectable and it is a strong
coolant in the outer cold regions of the disk. 
CO line luminosities are \s $10^{-9}$ L$_{\odot}$ in the $2\rightarrow 1$ line, and 
\s $10^{-7}$ L$_{\odot}$ in the higher $15\rightarrow 14$ transition for our
fiducial disk. Higher J transitions of CO in the far-infrared
and the sub-millimeter wavelength regions can be easily detected by Herschel
and can reveal the extent of the gas disk. 

Line ratios are also diagnostic of the gas characteristics. The [FeI]24$\mu$m line is
only strong when there is high shielding (or high columns and
densities) in the inner disk regions. This requires either high gas masses,
steep surface density distributions, small inner holes, or a combination of these. The
FeI/SI ratio is especially suited to probing these parameters. Significant amounts of 
[OI]63$\mu$m and CO (J= $15\rightarrow 14$) 187 $\mu$m radiation emanate from regions
$r\gtrsim5 $ AU, in contrast to the SI or \mh 17 $\mu$m lines. Therefore, the ratios
[OI]63$\mu$m/SI, [OI]63$\mu$m/\mh 17 $\mu$m,  CO 187 $\mu$m/\mh 17 $\mu$m are sensitive
to the relative masses of gas at $r\lesssim 5$ AU compared to that at 
$r\gtrsim5 $ AU. In other words, these ratios are
sensitive to the steepness of the surface density distribution. 

The overall strength of the mid-IR lines primarily depends on the mass of warm ($T\gtrsim 100$ K)
gas, which typically lies within 5 AU of the central star. Disks with more total gas mass,
with smaller inner holes, or with steeper surface density distributions, produce stronger
mid-IR emission. Because gas-dust collisions mainly heat the bulk of the gas, the
line emission also increases with either greater dust mass, or, holding dust mass
fixed, with smaller minimum grain size $a_{min}$. In the latter case, the smaller size
grains are warmer and have greater surface area than their larger counterparts, and 
therefore heat the gas more vigorously.

As little as  $10^{-3} M_J$ of warm gas at $r\lesssim5$ AU can be detected 
by the Spitzer Space Telescope in the [SI]25$\mu$m line for sources at \s 30 pc from the
Sun. However, to better diagnose the physical characteristics of the gas requires
the detection of several lines, and this requires roughly an order of magnitude
greater gas mass (\s $10^{-2}$ \mj) at this distance. We emphasize that the line
luminosities in the strongest lines ([SI], [SiII], [FeII] and [OI]) increase
very slowly or even decrease 
with increasing gas mass (see Fig.~\ref{vmass}). 
This is because the gas temperature in these disks is largely controlled
by the dust temperature, which is independent of gas mass,
and the gas emission lines at high gas mass become optically thick
so that their luminosities saturate.  Our results show that at the
distance of Taurus (140 pc), even 1 \mj\ of gas 
in intermediate-aged optically thin dust disks could be difficult to detect.
However, typical disks in Taurus are more likely to be younger 
and therefore abundant in small grains ($a_{min} \sim 50 $\AA). A
younger central star is also more luminous in FUV. This will lead to 
an increase in  both collisional gas-grain heating (due to
the larger surface area of small grains) and grain photoelectric heating
which will make the gas warmer and increase gas line emission significantly. 

Assuming that we know the distance to the source,
 the X-ray and FUV luminosities of the central star, and the dust mass
and dust distribution from the IR and sub-mm/millimeter continuum observations, we can
readily constrain the gas properties by comparing the
observed fluxes in various emission lines with our models. Ultimately, if the lines
are spectrally resolved by high resolution instruments on SOFIA or Herschel, the line
profiles will also help constrain the gas dynamics and spatial distribution. 

To conclude, we find exciting prospects for the detection by the Spitzer
Space Telescope, SOFIA or Herschel of infrared line emission from the gas in
0.3 AU to 20 AU regions of intermediate-aged disks
at distances $\lesssim$ 150 pc. The core accretion
model of giant planet formation strongly suggests an epoch ($t<10^{6.5}-10^7$
years) when the dust in the 1 AU$-$ 20AU region is relatively optically thin, due to
the growth of several earth-mass cores, but where $\gtrsim$ 1 \mj\ of gas persists
for an extended period. Therefore, the observations of line
emission coupled with the models presented here  will provide a measure of the
evolution of gas in disks as they transition from planet-forming
disks to the older, gas-free, debris disks and a test of the core-accretion
scenario for gas giant planet formation. 
The presence of even small amounts of gas at times $t \sim
10^7$ to $10^8$ years has important effects on planet formation in the 0.3-20 AU region.

\acknowledgements
We acknowledge very helpful discussions with all the members of the
Formation and Evolution of Planetary Systems (FEPS) Spitzer Legacy team,
especially John
Carpenter, Michael Meyer, Joan Najita, Deborah Padgett, Stu
Weidenshilling and Sebastian Wolf.  We also acknowledge helpful advice
from Al Glassgold and Inga Kamp.   Support for this work comes
from the NASA FEPS SIRTF Legacy grant and a NASA Origins of Solar
Systems grant.

\clearpage

\appendix

\section{Chemistry}
We adopt gas phase abundances from Savage \& Sembach (1996) towards the
line of sight through a diffuse cloud to the star $\zeta$ Oph. The
abundances are listed in Table~\ref{abun}. Table~\ref{species} lists the atomic, ionic and
molecular species included in our chemical code. Reaction rates are
mostly from the UMIST'99 database, except for some reactions which
are listed in Table~\ref{rates}. Apart from those listed here, X-ray photoionization
rates are calculated as described in \S D.3, and reactions involving
PAHs and vibrationally-excited \mh are from TH85. The UV photoionization
rates of atoms and molecules are calculated using their photoabsorption
cross sections multiplied by the stellar flux
and integrated over frequency.

\section{X-ray Heating and Chemistry}
\subsection{X-ray Spectrum}
The stellar X-ray spectrum is assumed to be similar in form to the wTTS in
Figure 2 of Feigelson \& Montmerle (1999). A broken power law is fit to the
data for the counts G(in counts s$^{-1}$ keV$^{-1}$) within the energy range
0.5-10 keV, which then gives the luminosity in erg s$^{-1}$ keV$^{-1}$,
$L(E) = G(E) E$.
\begin{eqnarray}
 L(E) & = &  1.2 L_X E^{-1.75} ( E >2 {\rm keV})\\
        & = &  0.18 L_X E         ( E< 2 {\rm keV}) \\
\end{eqnarray}
where $L_X$ is the total X-ray luminosity of the star in erg s$^{-1}$,
and the energy $E$ is expressed in keV. The X-ray energy flux is 
given by $F(E)=L(E)/4\pi D^2$, where $D=(r^2+z^2)^{1/2}$ is the distance 
from a point in the disk to the star. 
The heating due to X-rays is
\begin{equation}
H_X = \int_{0.5}^{10} F(E) e^{-\tau_X} \sigma(E) n_H f_h dE
{\rm \ erg\ cm}^{-3} {\rm \ s}^{-1}
\end{equation}
where $\sigma(E)$ is the total gas X-ray photoabsorption cross
section per H nucleus, $\tau_X$ is the X-ray optical depth given by  $\sigma(E)N_H$,
$N_H$ is the total column density of the gas towards the source,
$n_H$ is the number density and $f_h$ is the fraction of absorbed
energy that goes into heating the gas.
We also include secondary heating of X-rays due to
photoionization from the $n=2$ state of atomic hydrogen and follow exactly
the treatment of Shang et al.(2002).

\subsection{X-ray Photoabsorption Cross Section  $\sigma(E)$}
We use the model of Wilms et al. (2000) for the 
X-ray photoabsorption cross section per H nucleus. They obtain the total
photoionization cross section of the ISM by summing the cross sections
of the gas, molecules and grains. They use cosmic abundances for the
various elements, and add the contribution from grains. Our grains are
large and typical X-ray penetration depths are \s $0.1 \mu$m, and
therefore X-rays do not penetrate even our smallest dust grains. We use only the 
gas-phase abundances in our calculation of the cross-section, and
we adopt abundances from Savage \& Sembach (1996) through a diffuse
cloud towards the star $\zeta$ Oph. 
\normalsize
We correct Figure 1 of Wilms et al.(2000) to correspond to our depleted
abundances and then fit a power law to find the absorption cross section. We obtain
\begin{equation}
\sigma(E) = 1.2 \times 10^{-22} \left({{E}\over{1 {\rm keV}}}\right)^{-2.594}
\end{equation}
Compare to the fit to the Wilms et al.(2000) cross-section 
\begin{equation}
\sigma(E) = 2.27 \times 10^{-22} \left({{E}\over{1 {\rm keV}}}\right)^{-2.485}
\end{equation}

\subsection{X-ray Photoionization Rates of Atoms and Molecules}
The absorption cross section of a species $i$ for ionization is 
approximated to be of the functional form
\bd
\sigma_i(E) = \sigma_{0i} \left({{E}\over{1 {\rm keV}}}\right)^{\alpha}
\ed
We use the partial photoionization cross sections of Verner \& Yakovlev (1995)
and the fitting formulae they provide (Table 1 of their paper). These
formulae for the cross sections in the range
0.5-10 keV are then fit again with a power law of the
above form for each neutral atomic species. In some cases, the cross section is
fit with a broken power law within this range. Table~\ref{xray-rates} lists the values 
of $\sigma_{0i}$ in barns, $\alpha$ and the energy ranges in keV
where the fits are valid.
The primary rate of photoionization is given by the integral
of the photon number flux and the cross section
\bd
\zeta_i = 6.25 \times 10^{8} \int_{0.5}^{10} \sigma_i(E) {{F(E)}\over{E}} 
e^{-\tau_X(E)} dE {\rm \ s^{-1}}
\ed
where $F(E)$ is  in units of erg s$^{-1}$
 keV$^{-1}$, and the energy $E$ is in keV. The attenuation
factor $\tau_X$ has the same definition as before and is equal to
the product of the {\em total} photoabsorption cross section and the
column density towards the source.

Secondary ionizations are important for hydrogen and helium.
The fraction of primary electron energy
going into secondary ionization is (Shull \& van Steenberg 1985), 
\bd
\Phi_H = 0.391\left(1-{{13.6}\over{E}}^{0.41}\right)^{1.76} \approx 0.35
\ed
\bd
\Phi_{He} = 0.0554\left(1-{{24.6}\over{E}}^{0.46}\right)^{1.66} \approx 0.05
\ed
where $E$ is now in eV. The number of secondary photoionizations produced by a primary
electron of energy $E$ is equal to $E\Phi_H/13.6$ and $E\Phi_{He}/24.6$ 
for hydrogen and helium. The number of secondary ionizations by a
photon of energy $E$  is therefore $25.7 (E/1$keV$)$ for hydrogen and
$2.0 (E/1$keV$)$ for helium.
The total photoionization rate for hydrogen and helium are therefore,
\bd
\zeta_{T,H} =  6.25 \times 10^{8} \int_{0.5}^{10} \sigma_i(E) {{F(E)}\over{E}} 
e^{-\tau_X(E)} (1+25.7E) dE {\rm \ s^{-1}}
\ed
\bd
\zeta_{T,He} =  6.25 \times 10^{8} \int_{0.5}^{10} \sigma_i(E) {{F(E)}\over{E}} 
e^{-\tau_X(E)} (1+2.0E) dE {\rm \ s^{-1}}
\ed
where $E$ is in units of keV. 
We ignore secondary ionizations of all other species.
Photoionization rates of molecules are assumed to be the sum
of the rates of their constituent atoms, e.g., $\zeta_{OH} = 
\zeta_O + \zeta_H$ etc. These integrals are numerically evaluated for
each species with the expressions for the cross sections and the
X-ray flux, as a function of column $N_H$ towards the star.

\section{Gas Heating Mechanisms}
\paragraph{Gas-grain collisions}  Dust grains in the optically thin disk 
are heated by the stellar radiation field and collisional gas-grain 
energy transfer is often the dominant heating mechanism for the gas. 
The dust grain temperature is size-dependent, smaller grains are
less efficient at re-radiating stellar light and are thus
warmer than  larger grains. Collisions of gas molecules with dust grains 
can either heat or cool the gas, depending on the gas and dust grain
temperatures.  Our treatment is similar to that of Hollenbach \& McKee 
(1979, hereafter HM79) but adapted for
an arbitrary grain size distribution. The net heating (or cooling) due to collisions
is given by
\begin{equation}
\Gamma_1 = \alpha_T n_H v_{gas} \sum_{a_{min}}^{a_{max}} \pi a^2 n_{dust}(a) 
(2 k_B T_{dust}(a)- 2 k_B T_{gas}) {\rm \ ergs\ cm}^{-3} {\rm \ s}^{-1}
\label{hcoll}
\end{equation}  
where $\alpha_T \approx 0.3$ is the thermal accommodation coefficient (Burke \& Hollenbach 1983),
$n_H$ the gas hydrogen nucleus number density, $v_{gas}$  the thermal velocity of the gas,
$a$  the dust grain radius, that ranges from $a_{min}$ to
$a_{max}$,  $n_{dust}(a)$  the number density of particles of size $a$,
and $T_{dust}(a)$ and $T_{gas}$ are the temperatures of a dust grain of
size $a$ and the gas respectively. 

\paragraph{FUV pumping of \mh and collisional de-excitation}
\mh molecules absorb FUV photons from the star and from the ISRF and
are excited to a higher vibrational state. Subsequent de-excitation
by collisions with H atoms and ground state \mh
molecules deposits the excess energy as heat into the gas. 
\mh molecules can also be directly photodissociated
from the vibrationally excited state, which again leads
to heating. (See TH85 for a detailed discussion). The heating due
to FUV pumping and  collisional de-excitation is given by
\begin{eqnarray}
\Gamma_2 & = &  3.18 \times 10^{-12} n({\rm H}^*_2)
(10^{-12} T^{0.5} \exp(-1000/T) n(H) + \\ \nonumber  &&  1.4 \times 10^{-12}
T^{0.5} \exp(-18100/(T+1200)) n({\rm H}_2)) + \\ \nonumber &&  G_{11.26} (
1.36 \times 10^{-23} (n({\rm H}_2)\beta_1(\tau) 
 n({\rm H}^*_2)) \beta_2(\tau)) + \\ \nonumber &&  5.8 \times
10^{-24}  n({\rm H}^*_2)) {\rm \ ergs\ cm}^{-3} {\rm \ s}^{-1}
\end{eqnarray}
where $ n({\rm H}_2)$ and $ n({\rm H}^*_2)$ are the number densities of
\mh and vibrationally excited \mh respectively, 
$\beta_1(\tau)$ and $\beta_2(\tau)$  the corresponding self-shielding factors and $G_{11.26}$ is the
local photon energy flux  in the range $11.26-13.6eV$ (from the star and the ISRF,
attenuated by dust and other gas species), normalized to the Habing flux in the same band 
($2.26 \times 10^{-4}$ erg cm$^{-2}$ s$^{-1}$).
We use self-shielding factors for CO from van Dishoeck \& Black (1988)
and for \mh from Draine \& Bertoldi (1996). 

\paragraph{X-ray heating}
X-rays incident on gas photoionize atoms and molecules, and the 
primary electrons thereby generated are energetic enough to cause
multiple secondary ionizations. The heating term arises from the
thermalization of some of the energy of these secondary electrons, 
and is proportional to the ionization rate due to X-rays. The 
heating due to X-rays is given by
\begin{equation}
\Gamma_3 = \int_{E_{min}}^{E_{max}} F(E) e^{-N_H \sigma(E)}
 \sigma(E) n_H f_h dE
 {\rm \ ergs\ cm}^{-3} {\rm \ s}^{-1}
\end{equation}
where $F(E)$ is the X-ray photon flux, $N_H$ is the hydrogen nucleus
column density of gas towards the source, $\sigma(E)$ is the X-ray
absorption cross section of the gas per hydrogen nucleus
(see Appendix B), and
$f_h$ is the fraction of energy that is converted to heat, equal
to 0.1 for atomic gas and 0.4 for molecular gas. 
(See Maloney et al. 1996 for details.) We also include indirect
heating by X-rays due to atomic hydrogen ionization, 
the recombining H$^{+}$ resulting in a substantial $n=2$ population 
of H because of the trapping of Lyman $\alpha$ photons, followed by
photoionization or collisional de-excitation 
of $n=2$ hydrogen atoms. (See  Shang et al. 2002.) 
We ignore X-ray scattering from the surface layer into the disk, 
as we estimate the contribution from this process to be negligible. 

\paragraph{Grain photoelectric heating}
Energetic electrons emitted from dust grains due to the absorption of 
stellar FUV photons can collide with and heat the gas (Bakes \&
Tielens 1994). The efficiency of this process is dependent on the 
size of the dust grain and decreases with increasing grain size. As we consider 
large dust grains in our disk models,
this process is not as efficient as in the ISM where small
submicron sized dust grains are abundant. The grain photoelectric heating
term has been revised to include the effects of a cooler
radiation field (stars of lower spectral type) and increased grain
sizes. We  use corrected grain photoelectric efficiencies from
Weingartner \& Draine (2001) and  Weingartner  (2002, private communication) for
our calculations of grain photoelectric heating in disks. 
\begin{equation}
\Gamma_4 =   G_0 \sum_{a_{min}}^{a_{max}} n_{dust}(a) \pi a^2 \epsilon(T_{FUV},a) 
{\rm \ ergs\ cm}^{-3} {\rm \ s}^{-1}
\end{equation}
where $\epsilon$ is the photoelectric heating efficiency and is a function
of the grain size $a$, the equivalent blackbody temperature of the stellar FUV flux 
$T_{FUV}$, and $G_0$ and
$n_e$ (which determine the grain charge). The FUV flux spectrum from the star
(Figure~\ref{stellar-flux}) has a characteristic temperature $T_{FUV}\sim 10,000$ K.
$G_0$ is the local stellar flux between 6.0$-$13.6 eV, normalized to the Habing
flux. A similar term for the ISRF is also included, where $T_{FUV}$ 
is replaced by the temperature of the ISRF, 30000 K, although this contribution
is generally negligible. 

\paragraph{Drift heating}
Dust grains and gas molecules are subject to different forces in the
disks, with radiation pressure acting only on dust and gas being
constrained by thermal pressure gradients. They could have different
disk velocities and orbits, and this causes a viscous heating term
which could be important in dense disks (Kamp \& van Zadelhoff 2001,
Takeuchi and Artymowicz 2002). We
add drift heating to our disk models, though we do not solve for the
disk orbital dynamics of the dust or gas. The heating due to drift
is given by 
\begin{equation}
\Gamma_5 = 0.5 \mu_H n_H \sum_{a_{min}}^{a_{max}} n_{dust}(a) \pi a^2 
v^3_{drift}(a) {\rm \ ergs\ cm}^{-3} {\rm \ s}^{-1}
\end{equation}
where $v_{drift}(a)$ is the drift velocity between  gas and a dust
particle of size $a$. We use the equations of Takeuchi and Artymowicz (2002) for the
calculation of the drift velocity. 

\paragraph{Heating due to formation of \mh}
We include heating due to the formation of \mh in our model.
The formation of \mh either on grain surfaces or by reactions of
H atoms with H$^{-}$ leads to heating of the gas. The grain formation
mechanism provides \s 0.2 eV of kinetic energy going into the newly
formed \mh molecule, and \s 4.2 eV of energy initially deposited
as rovibrational internal energy of the molecule (Hollenbach, Werner
\& Salpeter 1971). The reaction with  H$^{-}$ is an exothermic
reaction which liberates \s 3.53 eV of energy and we assume that most of this
energy is initially in  rovibrational internal energy of the molecule.
Generally in disks, the gas densities are much greater than the
critical density $n_{cr}$ to de-excite the vibrationally excited \mh
molecule. Therefore, the rovibrational energy is quickly converted into heat. 
 Grain formation of \mh is not included in our standard runs,
and this heating mechanism is only included for the case where
we allow formation of \mh on grains. Heating due to formation of \mh
by H$^{-}$ is given by
\begin{equation}
\Gamma_6 = 7.34 \times 10^{-21}{ {{n_H}^2 x(H^{-}) x(H)} 
\over{1+n_{cr}/n_H}} 
{\rm \ ergs\ cm}^{-3} {\rm \ s}^{-1}
\end{equation}
where
\begin{equation}
n_{cr} = {{10^6 T^{-1/2}}\over{1.6 x(H) \exp[-(400/T)^2]+
1.4 x(H_2) \exp[-12000/(T+1200)]}} {\rm cm}^{-3}
\end{equation}
and $x(H), x(H^{-})$, and $x(H_2)$ are the abundances of 
H, H$^{-}$ and \mh respectively (HM79). 
Heating due to formation of \mh on dust grains is given by
\begin{equation}
\Gamma_7 = 4.8 \times 10^{-29} {n_H}^2 f_{dust} x(H) 
(0.2 + {{4.2}\over{1+n_{cr}/n_H}}) 
{\rm \ ergs\ cm}^{-3} {\rm \ s}^{-1}
\end{equation}
and $f_{dust}$ is a factor that accounts for the change in dust 
surface area per H nucleus caused by changes in the 
size distribution and the gas-to-dust mass ratio as compared to 
interstellar values. 

\paragraph{Cosmic ray heating and other heating mechanisms}
Apart from the processes discussed  above, we consider
cosmic ray heating ($\Gamma_8$), photoionization of neutral carbon
($\Gamma_9$, see TH85), and heating by the excitation of [OI]63$\mu$m upper
level by the IR continuum radiation field followed by the
 collisional de-excitation of this level ($\Gamma_{10}$,
see Appendix D).  
\begin{equation}
\Gamma_8 = 1.5 \times 10^{-11}  \zeta n_H (x(H)+ 2 x(H_2))
{\rm \ ergs\ cm}^{-3} {\rm \ s}^{-1}
\end{equation}
where $\zeta$ is the primary cosmic ray ionization rate. 
\begin{equation}
\Gamma_9 = 4.16 \times 10^{-29} n_H x(C) G_{11.26} \exp(-3 A_v-\tau_{11.26}) \beta_C
{\rm \ ergs\ cm}^{-3} {\rm \ s}^{-1}
\end{equation}
where  $x(C)$ is the abundance of neutral carbon, $\tau_{11.26}$ is the 
optical depth due to gas absorption towards the star in the FUV band 
that photoionizes carbon,
and $\beta_C$ is a self-shielding factor (van Dishoeck \& Black 1988). To this
term we also add similar terms for the contribution from the ISRF, and use
optical depths and self-shielding factors due to gas columns vertically up
and down from the spatial position in the disk. 

\section{Gas Cooling Processes}
Gas cooling occurs by gas-grain collisions when the gas is warmer than
the dust (Eq.~\ref{hcoll}) and by collisional excitation followed by
line radiation of gaseous atoms, ions and molecules. We essentially follow the 
treatment of TH85, in which the cooling due to a transition from 
level $i$ to level $j$ is given by
\begin{equation}
\Lambda_{ij} = n_i A_{ij} h \nu_{ij} \beta_{esc}(\tau_{ij})
\left(1-P( \nu_{ij} )/S( \nu_{ij})\right),
\end{equation}
where $n_i$ is the population density of the species in level $i$,
$A_{ij}$ is the spontaneous transition probability, $ h \nu_{ij}$ is the
energy of the emitted photon, $\tau_{ij}$ is the optical
depth in the line and $ \beta_{esc}(\tau_{ij})$ is the 
associated escape probability. $P( \nu_{ij}) $ is the background
radiation term which consists of the sum of the cosmic microwave 
background radiation and the infrared emission from the dust grains of 
different temperatures at the wavelength of the transition. The 
source function $S( \nu_{ij})$ is determined by solving for the
level populations assuming statistical equilibrium. We generally assume that
radiation can escape from any point in the disk either vertically up
or down or radially out the inner radius.
 We calculate the column densities in each level of the
cooling species in all three directions. From the column densities  we
 determine the optical depth in the line and the escape probability
of a photon, which then gives the cooling rate in a given line.
If the gas temperature is  lower than that of dust, and
the background radiation  stronger than the source function, 
the transition can  heat the gas as the term 
$1-P( \nu_{ij} )/S( \nu_{ij})$ becomes negative. In this case, a background
photon is absorbed by a particular species and subsequent collisional
de-excitation transfers the excitation energy to gas heating. 

We treat CO and \mh cooling in a detailed multi-level calculation
but treat OH and H$_2$O somewhat differently. 
Analytical approximations to multi-level line cooling calculations of
OH and H$_2$O are derived for use in the gas disk code. The full level
population calculation was prohibitively expensive to include in  code
and instead,  we derived analytic fits for the total cooling similar to HM79. 
For H$_2$O, we used collisional rate data for approximately 170 levels  
for both the ortho and para species, and new parameters (refer HM79)
that best fit the data, $\sigma=2.0 \times 10^{-15}$ cm$^{-2}$, E$_0/K$=23.0 K and
A$_0 = 6.0 \times 10^{-3}$ s$^{-1}$. The collisional rate data for OH for the
first 45 levels are used to obtain the values, $\sigma=8.0 \times 10^{-16}$ cm$^{-2}$, 
E$_0/K$=5.4 K and A$_0 = 7.6 \times 10^{-4}$  s$^{-1}$. 
\begin{table}[f]
\caption{ Elemental gas phase abundances (from Savage \& Sembach 1996)}
\label{abun}
\begin{tabular}{lr}
\hline
Element & Abundance \\
\hline
He & 0.1 \\
C & 1.4 $\times 10^{-4}$ \\
O & 3.2 $\times 10^{-4}$ \\
Si & 1.7 $\times 10^{-6}$ \\
Mg & 1.1 $\times 10^{-6}$ \\
Fe & 1.7 $\times 10^{-7}$ \\
S & 2.8 $\times 10^{-5}$ \\
\hline
\end{tabular}
\end{table}
\begin{table}[f]
\caption{Chemical species included in the models}
\label{species}
\begin {tabular}{lllllllll}
\hline
H & He & C & O & \mh & O$_2$ & OH & CO & H$_2$O \\
H$^{+}$ & He$^{+}$ & C$^{+}$ & O$^{+}$ & OH$^{+}$ & CO$^{+}$ &
H$_2$O$^{+}$ & HCO$^{+}$ & H$_3$O$^{+}$ \\
\mh$^{+}$ & H$_3$$^{+}$ & CH$^{+}$ & CH$_2$$^{+}$ & CH$_3$$^{+}$ &
CH & CH$_2$ & Mg & Mg$^{+}$ \\
Si & Si$^{+}$ & SiH$_2$$^{+}$ & SiH & SiO & Fe & Fe$^{+}$ & S &S$^{+}$ \\
SiO$^{+}$ & HOSi & H$^{-}$ & 
& CH$_4$&CH$_3$ 
CH$_4$$^{+}$ & HS$^{+}$ & H$_2$S$^{+}$ & H$_3$S$^{+}$ \\
HS & H$_2$S& CS & CS$^{+}$& HCS$^{+}$ 
HCS & OCS&OCS$^{+}$&HOCS$^{+}$ \\
SO&SO$^{+}$&HSO$^{+}$&SO$_2$&SO$_2$$^{+}$
HSO$_2$$^{+}$&H$_2$CS&H$_2$CS$^{+}$&H$_3$CS$^{+}$ \\
S$_2$$^{+}$& S$_2$H$^{+}$&\\
\hline
\end{tabular}
\end{table}
\begin{table}[f]
\caption{ Reaction rates where they differ from the UMIST'99 database.}
\label{rates}
\begin{tabular}{lclclllclclr}
\hline
H$_3$O$^+$ & $+$ & e$^{-}$ & & & $\rightarrow$ &  H$_2$O & $+$  & H &  & & 
 $4.0 \times 10^{-6} T^{-1/2}$ \\
H$_3$O$^+$ & $+$ & e$^{-}$ & & & $\rightarrow$ &  H$_2$  & $+$ & H & $+$ & O &
 $1.73 \times 10^{-7} T^{-1/2}$ \\
H$_3$O$^+$ & $+$ & e$^{-}$ & & & $\rightarrow$ &  H$_2$  & $+$ & OH &  &  &
 $3.3 \times 10^{-6} T^{-1/2}$ \\
H$_3$O$^+$ & $+$ & e$^{-}$ & & & $\rightarrow$ &  OH & $+$ & H & $+$ & H &
 $9.5 \times 10^{-6} T^{-1/2}$ \\
H$_3$$^+$  & $+$ & e$^{-}$ & & & $\rightarrow$ &  H$_2$ & $+$ & H & & & 
 $2.9 \times 10^{-7} T^{-1/2}$ \\
H$_3$$^+$  & $+$ & e$^{-}$ & & & $\rightarrow$ &  H & $+$ & H & $+$ &  H &
 $9.0 \times 10^{-7} T^{-1/2}$ \\
H & $+$ & H &$+$ & H &  $\rightarrow$ &  H$_2$ & $+$ & H & & & 
 $8.8 \times 10^{-33}$ \\
H$_2$ &$+$&  H & $+$&  H & $\rightarrow$ &  H$_2$ & $+$& H$_2$ & & &
 $2.8 \times 10^{-31} T ^{-0.6} $ \\
\hline
\end{tabular}
\end{table}
\begin{table}[f]
\caption{Fits to photoionization cross sections where the cross section is 
given by $\sigma_i(E) = \sigma_{0i}(E/1keV)^{\alpha}$ and $E_1<E<E_2$.}
\label{xray-rates}
\begin{tabular}{ccccc}
\hline
Atom & $\sigma_{0i}$ barns & $\alpha$ & $E_1$ keV  & $E_2$ keV\\
\hline
H & 11.55 & -3.4 & 0.5 & 10 \\
He & 43.12 & -3.3 & 0.5 & 10 \\
C & 4.56 $\times 10^{4}$ & -3.01 &  0.5 & 10 \\
O & 7.34 $\times 10^{3}$ & -3.125 &  0.5 & 0.537 \\
  & 1.28 $\times 10^{5}$ & -2.9 &  0.537 & 10 \\
Mg & 5.93 $\times 10^{4}$ & -2.38 &  0.5 & 1.312 \\
 & 5.5 $\times 10^{5}$ & -2.76 &  1.312 & 10 \\
Si & 1.00 $\times 10^{5}$ & -2.44 &  0.5 & 1.835 \\
 & 9.12 $\times 10^{5}$ & -2.74 &  1.835 & 10 \\
S & 1.686 $\times 10^{5}$ & -2.485 &  0.5 & 2.477 \\
 & 1.39 $\times 10^{6}$ & -2.7 &  2.477& 10 \\
Fe & 1.63 $\times 10^{5}$ & -1.98 &  0.5 & 0.726 \\
 & 2.075 $\times 10^{6}$ & -2.0 &  0.726 & 0.851 \\
 & 1.05 $\times 10^{6}$ & -2.52 &  0.851 & 7 \\
 & 6.67 $\times 10^{6}$ & -2.627 &  7 & 10 \\
\hline
\end{tabular}
\end{table}

\clearpage
\clearpage

\clearpage

\clearpage

\clearpage

\clearpage

\clearpage

\clearpage

\clearpage

\clearpage

\clearpage

\clearpage

\begin{figure}[f]
\plotone{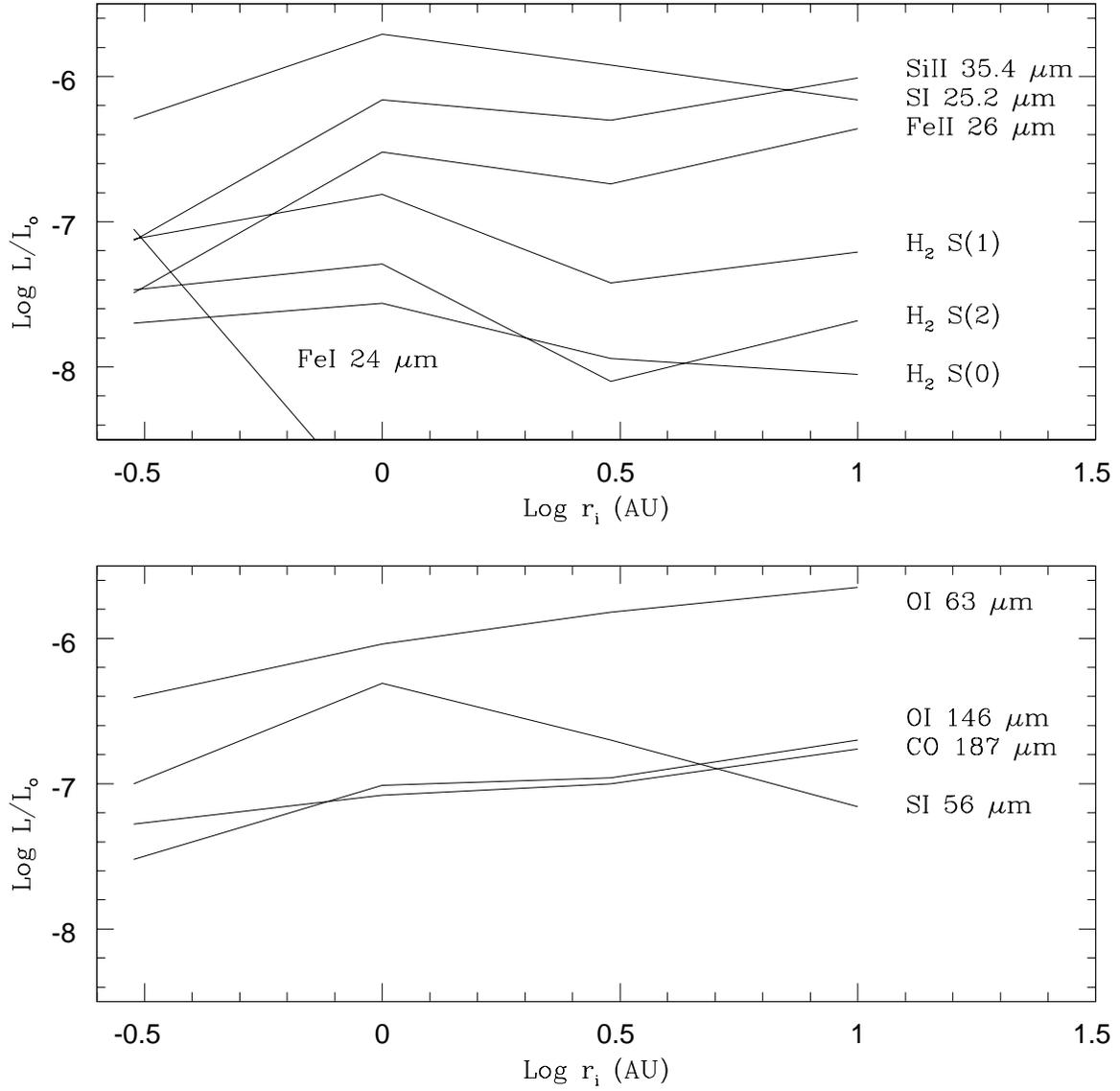}
\caption{Gas line luminosities for disk models with differing inner
radii, $r_{i}$, and other parameters as in the standard case.}
\label{vrmin}
\end{figure}
\clearpage

\begin{figure}[f]
\plotone{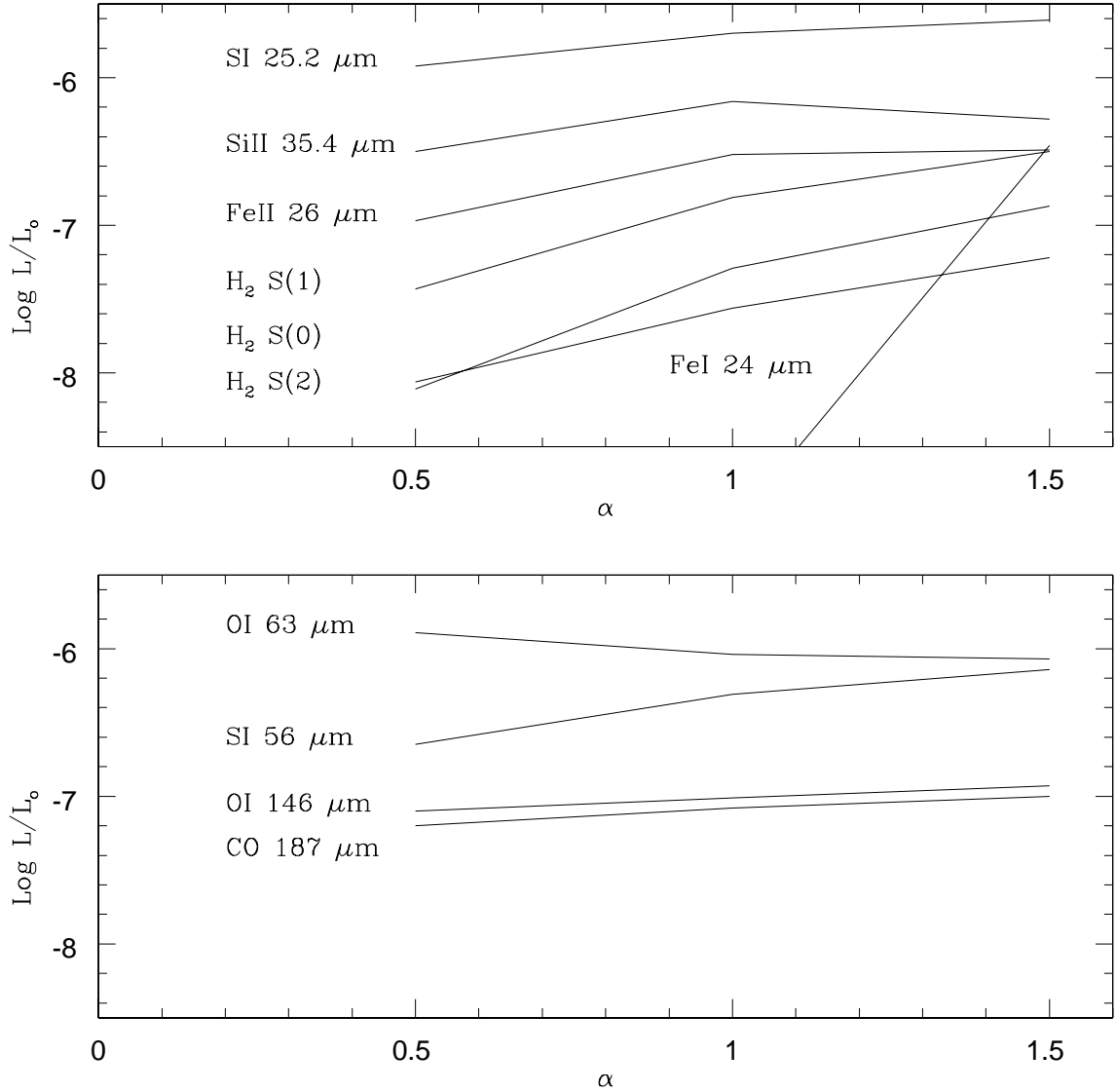}
\caption{Changes in the line luminosities due to variations in the
slope of the power law index $\alpha$ of the surface density distribution.}
\label{valpha}
\end{figure}
\clearpage

\begin{figure}[f]
\plotone{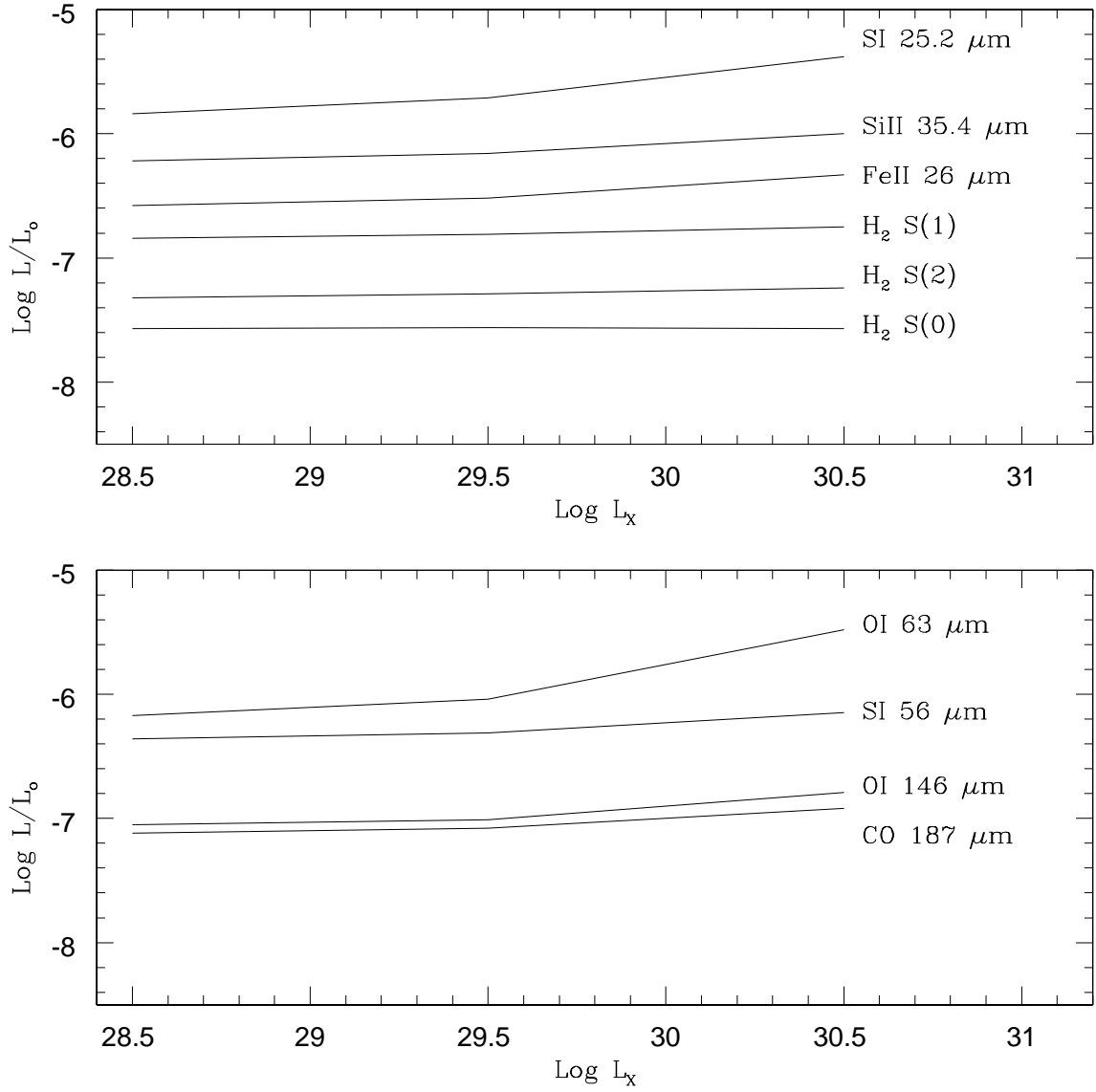}
\caption{Gas line luminosities for disks around a G star with
different X-ray luminosity.}
\label{vLx}
\end{figure} 
\clearpage

\begin{figure}[f]
\plotone{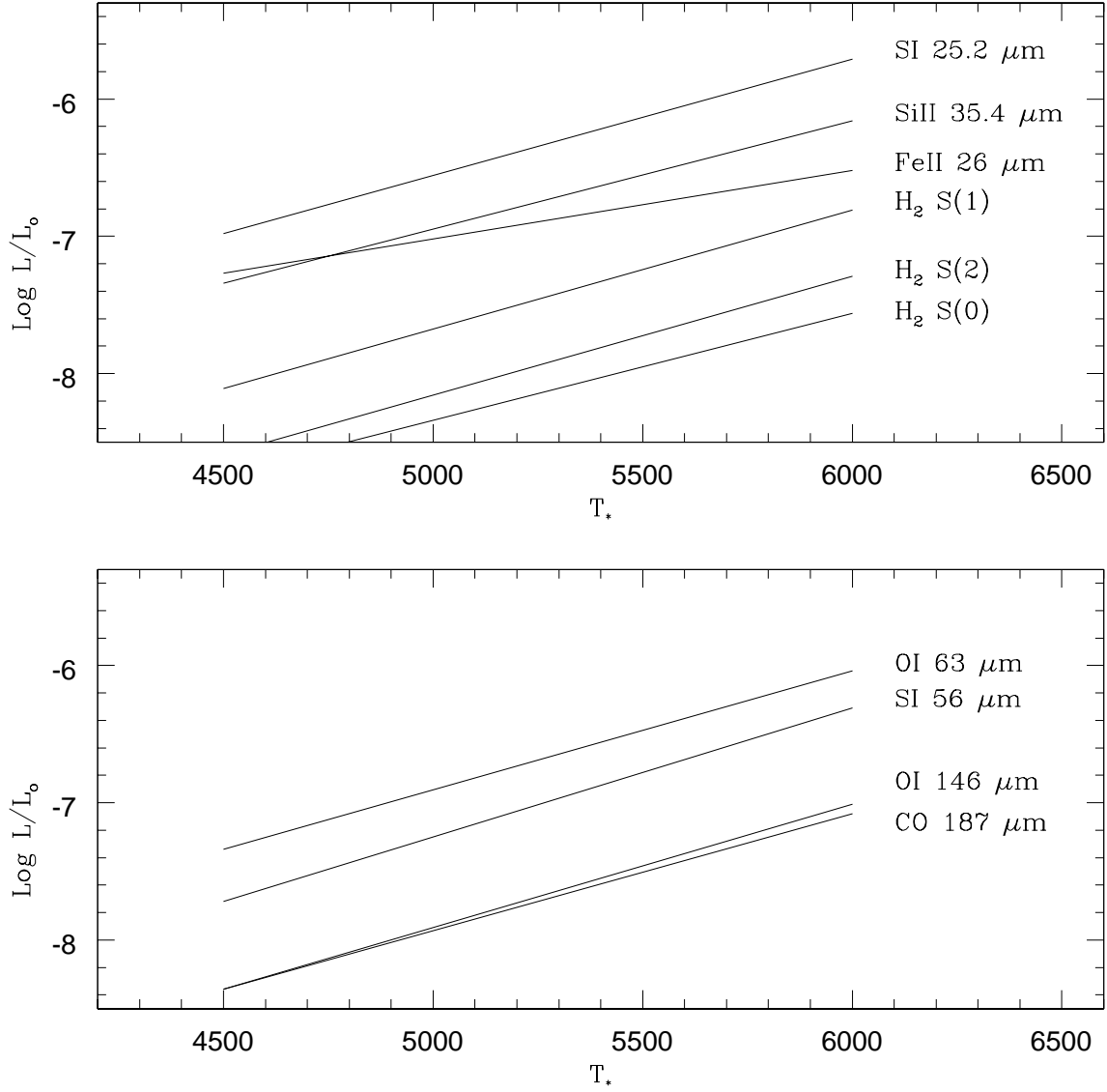}
\caption{Gas line luminosities for disks with parameters as in
the standard disk model but  around a G star($T_*=6000$) and a K star($T_*=4500$).}
\label{vstar}
\end{figure}

\begin{figure}[f]
\plotone{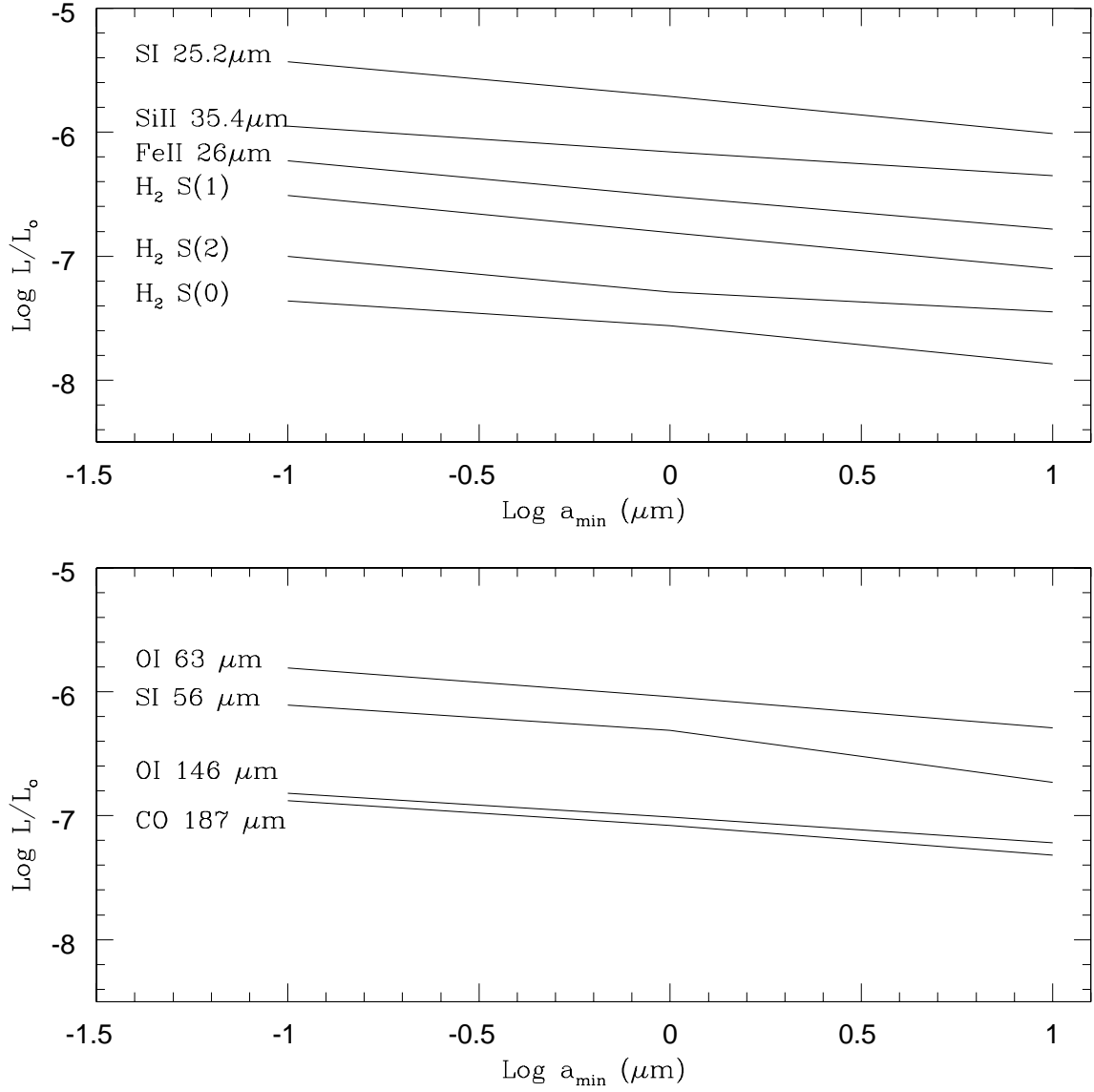}
\caption{Gas line luminosities for disks with dust distributions where the
minimum grain size $a_{min}$ is varied. }
\label{vamin}
\end{figure} 
\clearpage

\end{document}